\documentclass[10pt,journal,compsoc]{IEEEtran}
\ifCLASSOPTIONcompsoc
  \usepackage{cite}
\else
  \usepackage[nocompress]{cite}
\fi

\ifCLASSINFOpdf
\else
\fi

\usepackage{tablefootnote}
\usepackage{tabularx}
\usepackage[T1]{fontenc}
\usepackage[utf8]{inputenc}
\usepackage{csquotes} 
\usepackage[font=footnotesize,labelfont=bf]{caption}
\usepackage[font=footnotesize,labelfont=bf]{subcaption}
\usepackage{graphicx}
\usepackage{amsmath,amssymb,amsfonts}
\usepackage{algorithm,algorithmicx,algpseudocode}
\algrenewcommand\algorithmicrequire{\textbf{Input:}}
\algrenewcommand\algorithmicensure{\textbf{Output:}}
\usepackage{url}
\usepackage{hyperref}
\usepackage{array}
\usepackage{kotex}
\usepackage{color, colortbl}
\usepackage{longtable}
\usepackage{xcolor}
\usepackage{ragged2e}

\newcolumntype{L}[1]{>{\raggedright\let\newline\\\arraybackslash\hspace{0pt}}m{#1}}
\newcolumntype{C}[1]{>{\centering\let\newline\\\arraybackslash\hspace{0pt}}m{#1}}
\newcolumntype{R}[1]{>{\raggedleft\let\newline\\\arraybackslash\hspace{0pt}}m{#1}}
\usepackage{enumerate}
\usepackage{tikz}
\usetikzlibrary{fadings}
\usetikzlibrary{arrows.meta}
\tikzset{%
  >={Latex[width=2mm,length=2mm]},
            base/.style = {rectangle,  draw=black,
                           minimum width=2.2cm, minimum height=1cm,
                           text centered, font=\sffamily},
  activityStarts/.style = {base, fill=blue!20, middle color=white},
         process/.style = {base, minimum width=2.2cm, fill=blue!20, middle color=white,
                           font=\ttfamily},
}

\begin{document}
%
\title{Provenance-enabled Packet Path Tracing in the RPL-based Internet of Things }

 \author{Sabah~Suhail,
        Mohammad~Abdellatif,
         Shashi~Raj~Pandey,
         Abid~Khan,
         and~Choong~Seon~Hong,~\IEEEmembership{Senior~Member,~IEEE}
         
\thanks{S. Suhail, S. R. Pandey, and C. S. Hong  are with the Department of Computer Science and Engineering, Kyung Hee University, Yongin 446-701, South Korea (e-mail: sabah@khu.ac.kr; shashiraj@khu.ac.kr; cshong@khu.ac.kr).}
\thanks{M. Abdellatif is with the Department of Electrical Engineering, British University in Egypt Cairo, Egypt (e-mail: mohammad.abdellatif@bue.edu.eg).}
\thanks{A. Khan is with the Department of Computer Science, COMSATS, Islamabad, Pakistan (e-mail: abidkhan@comsats.edu.pk).}
}
\IEEEtitleabstractindextext{%
\justify
\begin{abstract}
The interconnection of resource-constrained and globally accessible things with untrusted and unreliable Internet make them vulnerable to attacks including data forging, false data injection, and packet drop that affects applications with critical decision-making processes. For data trustworthiness, reliance on provenance is considered to be an effective mechanism that tracks both data acquisition and data transmission. However, provenance management for sensor networks introduces several challenges, such as low energy, bandwidth consumption, and efficient storage. This paper attempts to identify packet drop (either maliciously or due to network disruptions) and detect faulty or misbehaving nodes in the Routing Protocol for Low-Power and Lossy Networks (RPL) by following a bi-fold provenance-enabled packed path tracing (PPPT) approach. Firstly, a system-level ordered-provenance information encapsulates the data generating nodes and the forwarding nodes in the data packet. Secondly, to closely monitor the dropped packets, a node-level provenance in the form of the packet sequence number is enclosed as a routing entry in the routing table of each participating node. Lossless in nature, both approaches conserve the provenance size satisfying processing and storage requirements of IoT devices. Finally, we evaluate the efficacy of the proposed scheme with respect to provenance size, provenance generation time, and energy consumption.
\end{abstract}

}

\maketitle

\IEEEdisplaynontitleabstractindextext

%
\IEEEpeerreviewmaketitle

\IEEEraisesectionheading{\section{Introduction}\label{sec:introduction}}

\IEEEPARstart{T}{he} IPv6 over Low-power Wireless Personal Area Network (6LoWPAN)~\cite{thubert2011compression,kushalnagar2007ipv6} has enabled the interconnection of \textit{things} and the Internet, thus forming the Internet of Things (IoT). Considering the connectivity between low-power and lossy 6LoWPAN networks with resource-constrained things, Routing Protocol for Low-Power and Lossy Networks (RPL) has been standardized as one of the candidate protocols by the Internet Engineering Task Force (IETF) routing over low power and lossy network (ROLL) group to address the issues including scarce resources, lossy link characteristics and different routing requirements \cite{winter2012rpl}. RPL is a proactive distance vector protocol for low power and lossy networks (LLNs) that operates by establishing a directed acyclic graph (DAG) based on a set of routing metrics and constraints. Despite having the cutting-edge crypto-based solutions to provide data security, the self-configuring sensor nodes operated by RPL are susceptible to various attacks \cite{shreenivas2017intrusion} especially due to the global accessibility of unattended things.

By default, the RPL implementations do not enable secure operation modes. Therefore, malicious nodes can make the network vulnerable to multiple attacks \cite{glissa2016secure}. For instance, attacks on data (e.g., data forging, false data injection, data fabrication, and data replay), and attacks causing network disruptions (e.g., resource depletion, network congestion, packet loss, link quality degradation, and hole creation). In addition to these attacks, the untrustworthy data traversal across unattended sensor nodes may cause issues during data debugging, data verification and data authenticity \cite{suhail2016introducing}.

Data trustworthiness is an important requirement especially for applications relying on sensor data for critical decision-making processes, risk assessment, and performance evaluation. To enforce trustworthy data in transit and routing with quality of services (QoS), it is essential to keep track of the complete data propagation path. Such tracking of data can be formulated through \textit{Provenance}. Provenance is a meta-data describing the complete lineage of data and a set of actions performed on data \cite{hasan2009case}. Provenance ensures the integrity of data during data debugging, reconciliation, replication, decision making, performance tuning, auditing, and forensic analysis \cite{zafar2017trustworthy}.

Provenance has been extensively studied in a variety of application areas including databases, scientific work-flows, distributed systems, and networks. It has many ingredients (for instance, confidentiality, integrity, availability, privacy, and non-repudiation) as discussed in \cite{zafar2017trustworthy}. In wireless sensor networks (WSNs), the idea of incorporating provenance has been discussed in \cite{shebaro2012demonstrating, sultana2013secure, sultana2015lightweight, DBLP:conf/wowmom/AlamF11, wang2016dictionary, xu2018cluster, hussain2014secure}. In comparison to WSNs, the idea of provenance in the IoT is yet to be explored keeping in view the challenging requirements and constraints associated with the novel architecture of the IoT. For instance, in the IoT, things are globally accessible, operating in an unattended environment, resource constrained in nature, and connected through the lossy links. Keeping in view the constraints associated with IoT and the challenges associated with provenance collection mechanism \cite{suhail2018data}, it is more feasible to take into account the recommended underlying protocol for IoT devices such as RPL. Such consideration allows the exploitation of existing infrastructure of the protocol, and hence, put less burden on the resource-consuming nodes during provenance collection and validation. In this regard, introducing provenance scheme for RPL-connected IoT devices is a viable option because RPL is a highly recommended protocol for sensor motes. 

In this paper, we have focused on formulating a provenance-based framework for RPL-based IoT networks by introducing \textit{Provenance-based Packet Path Tracing} (PPPT) scheme. The proposed scheme not only captures the data generating source nodes but also makes the intermediate nodes accountable for forwarding the packet to the destination, ensuring trustworthy and quality data in resource constrained things. The main contributions of this paper are summarized as follows.
\begin{itemize}
    \item We investigate the significance of integrating provenance in RPL-based IoT by highlighting the problem of providing data trustworthiness in sensor motes.
    \item With reference to the resource-constrained things, we propose a lossless provenance scheme that operates by acquiring \textit{system-level provenance} information regarding the path traversed by the packet. For \textit{node-level provenance}, we exploit the routing information in the routing table at intermediate nodes by including the packet sequence number in the routing table. The inclusion of both system-level and node-level provenance helps in detecting malicious nodes or configuration of other network discrepancies.
    \item To mitigate the need for processing the forged or missing provenance in the data packet and to speed up the process of data trust assessment, we set a bit in IPv6 extension header (EH). 
    \item We design a packet path provenance encoding and decoding mechanisms for RPL storing mode particularly for Multipoint to Point (M2P) scenario. We evaluate the proposed technique with respect to provenance size, provenance generation time, additional storage overhead in terms of RAM and ROM, robustness and energy consumption metrics.
    \item We provide a comparison in terms of provenance generation time and average power consumption between RPL-only and the proposed provenance-enabled RPL PPPT scheme.
    \item We also compare our approach with bloom filter (BF) based techniques \cite{sultana2013secure,sultana2015lightweight} and Provenance as ID (PID) \cite{suhail2018data} in terms of provenance size. 
\end{itemize} 
The rest of the paper is organized as follows. Section \ref{rpl} provides an overview of RPL protocol. Section \ref{IoTsys} discusses the significance for integration of provenance in the IoT and the associated challenges. Section \ref{SystemModel} introduces the system model. Section \ref{scheme} describes the provenance encoding and decoding algorithms. Section \ref{claims} discusses the security analysis of our approach. Section \ref{simulation} presents the simulation results. Section \ref{relatedwork} surveys related work, and finally, Section \ref{conclusion} concludes the paper with future work.
\begin{figure}[t!]
\centerline{\includegraphics[width=3.0in]{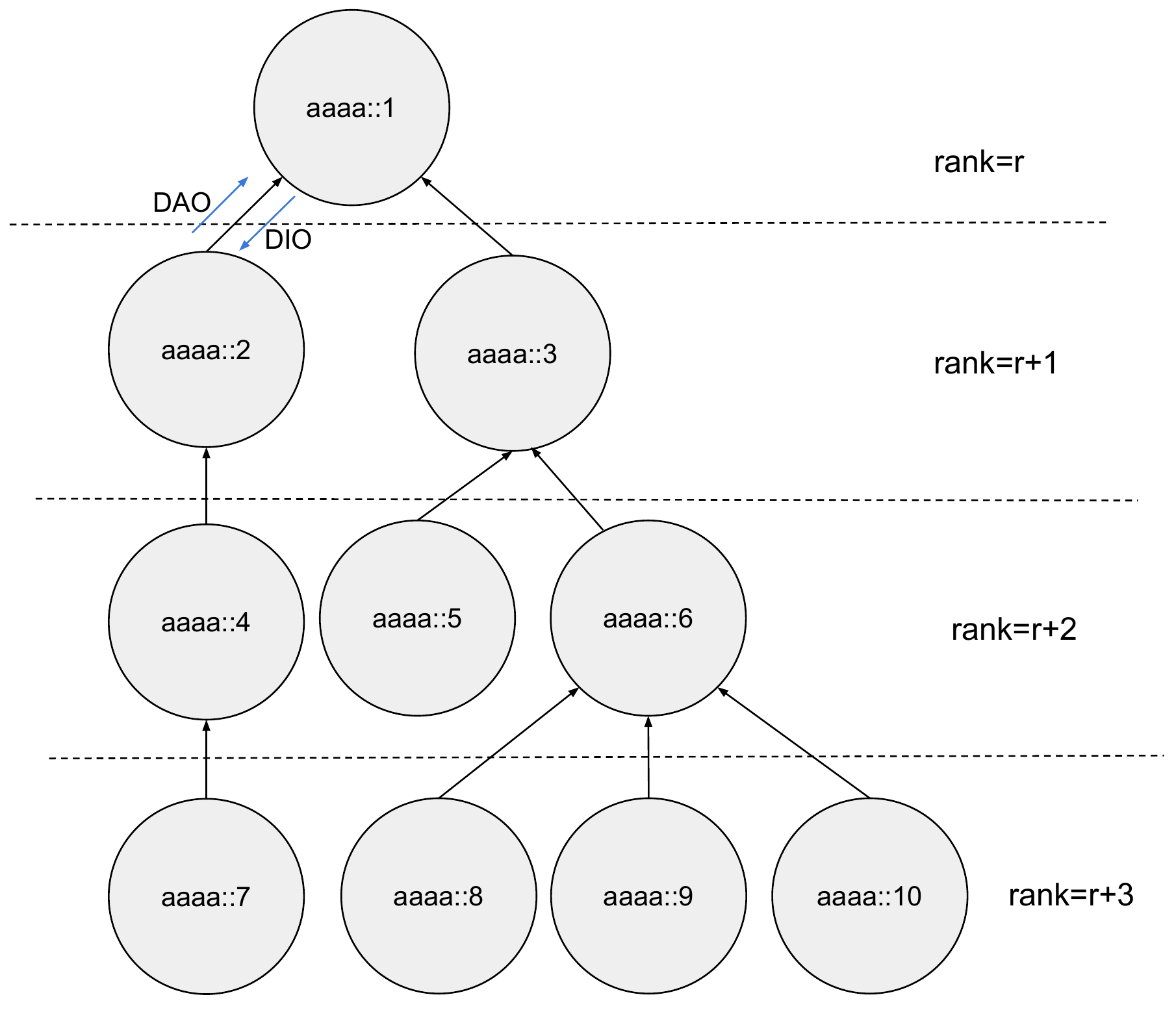}}
\caption{A sample RPL DODAG instance showing multipoint-to-point scenario. Each node has a unique IPv6 address. Each parent node maintains a neighboring table (NT) and routing table (RT). For example, node 3 keeps information about its neighboring nodes (1, 2, 5, 6) in NT and also routing entries (5, 6, 8, 9, 10) in RT.}
\label{fig:DODAG}
\end{figure}

\section{Overview of the RPL protocol} \label{rpl}
RPL is a standardized routing protocol for the IP-connected IoT. It creates a Destination-Oriented Directed Acyclic Graph (DODAG) and supports 2 modes of operation including uni-directional traffic from nodes to the root node (Multipoint-to-Point), and bi-directional traffic between nodes and the root node (Point-to-Multipoint and Point-to-Point). A typical RPL DODAG is shown in Fig. \ref{fig:DODAG} where each node has a node ID (an IPv6 address), one parent (preferred parent) or more parents (substitute parents), and a list of neighbors.
\subsection{DODAG Topology Formation} \label{rpl-topology}

RPL uses three types of control messages i.e., DODAG Information Object (DIO), DODAG Information Solicitation (DIS), and DODAG Destination Advertisement Object (DAO) for creating and maintaining RPL topology \cite{ali2012performance}. The RPL DODAG topology formation process is triggered by the root node by emitting DIO messages after neighbor discovery. The DIO message contains information about RPL Instance, DODAG, objective function (OF) and other configuration parameters. Upon the reception of DIO, the neighboring nodes compute their rank (i.e., the position of a node with respect to root node) based on the OF. The DIO message also informs about downward or upward routing based on Mode of Operation (MOP) flag.  
In case the downward routes are preferred, the child nodes would trigger the DAO messages in order to advertise their reachability in the DODAG towards their parents along with DAO lifetime and other parameters. A new node may join an existing network by broadcasting a DIS message in order to solicit a DIO message from a neighboring node. The control messages containing the routing information and other metrics disseminate information aperiodically, whereas the periodicity is determined through the trickle timer algorithm \cite{levis2011trickle}.

\subsection{Packet Routing}
In a DAG, a node may either act as a source node (also known as data generating node) or a forwarding node/aggregator node. In order to send a packet from source to destination (typically root node), the source node forwards the packet to its preferred parent that takes the forwarding decision based on the routing information in its routing table (RT) until it arrives at destination (root) node. The root usually performs the job of information retrieval, data processing or data aggregation. 

\section{Provenance-aware IoT system for resource-constrained nodes} \label{IoTsys}
\begin{figure}[t!]
\centerline{\includegraphics[width=2.5in]{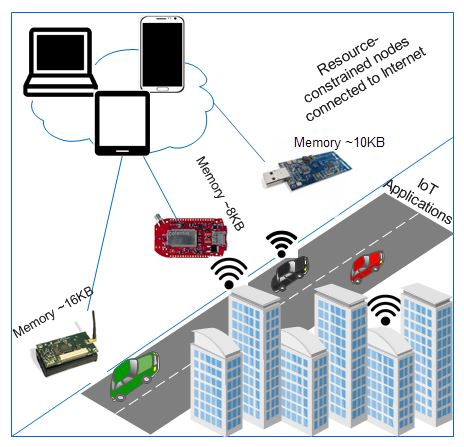}}
\caption{IoT application showing the interconnection of resource-constrained nodes to the Internet.}
\label{fig:IoTapp}
\end{figure}

The IoT is gaining ground as a pervasive presence around us comprising of a variety of things or objects – such as RFID tags, sensors, actuators, mobile phones, and others that are able to interact with each other or with their neighbor nodes through unique addressing schemes to reach common goals \cite{atzori2010internet}. Fig. \ref{fig:IoTapp} shows the interconnection of resource-constrained nodes with the Internet. The things composing the IoT are limited by low resources in terms of both computation and energy capacity and therefore, they cannot implement complex schemes supporting security. The deployed sensor motes in a smart city scenario (as shown in Fig. \ref{fig:IoTapp}) have low storage capacity ranging from 8KB to 16KB. Furthermore, IoT is extremely vulnerable to attacks because (i) they operate in unattended domain, (ii) eavesdropping and data loss in lossy wireless communication, (iii) components are characterized by low capabilities in terms of both energy and computing resources. Hence, unquestionably the data from sensor nodes require trustworthiness that can be achieved through provenance. Integration of provenance in resource-consuming nodes is a challenging task because of limited communication data rate, memory, energy, and processing constraints. Employing a forensic-aware system as data provenance in these resource-hungry devices to establish the authenticity of data entails a contemplation. In this section, we discuss the need for introducing provenance in the IoT. We also investigate the challenges associated with IoT-aware provenance system. 
\begin{figure}[t!]
\centerline{\includegraphics[width=2.5in]{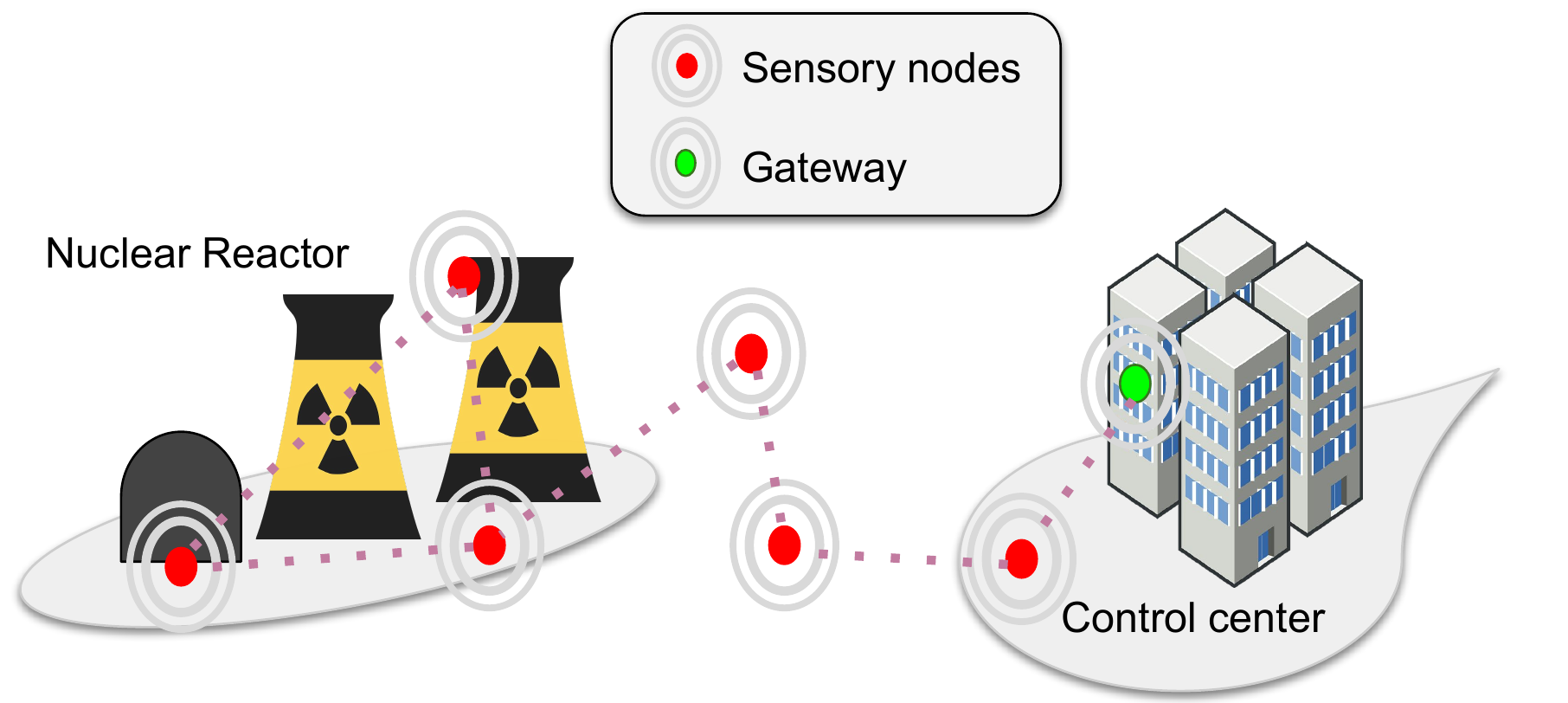}}
\caption{Nuclear reactor scenario showing sensory nodes relaying data to the control center.}
\label{fig:reactor}
\end{figure}
\subsection{Motivation: Integration of Provenance in the IoT}
Cost effective miniaturized devices with computational and networking capabilities are providing promising solutions to optimize the comfort and quality life style in various pervasive application areas including domotics, smart cities, smart grid, environment monitoring, meteorology, energy management, health-care systems, industrial automation, and precision agriculture. According to 
US National Intelligence Council (NIC), by 2025 Internet nodes may reside in everyday things such as food packages, furniture, paper documents, and more \cite{USReport}. Despite these benefits, the revolutionary technology of IoT is vulnerable to security issues causing user dissatisfaction (for instance, random bugs), privacy violation (for instance, eavesdropping\footnote{Data from connected CloudPets teddy bears leaked and ransomed exposing kid's voice messages, CNET, February 2017.})\footnote{Spying kettles and electric irons, BBC, November 2013.}, monetary loss (for instance, denial-of-service attacks or ransom-ware\footnote{Hospitals across the UK hit by WannaCrypt ransom-ware cyberattack, ZDNet, May 2017.}), or even loss of life (for instance, attackers controlling vehicles\footnote{Hackers Remotely Kill a Jeep on the Highway with Me in It, WIRED, July 2015.}, triggering malicious instructions to jeopardize a patient's life by hacking of Medtronic insulin pump\footnote{Medtronic probes insulin pump risks, Reuters, October 2011.} to dose fatal amount of insulin to diabetes patients, and hacking implantable cardiac devices including  pacemakers and defibrillators to deplete the battery or administer incorrect pacing or shocks\footnote{Vulnerabilities found in St.Jude Medical implantable cardiac devices, ZDNet, September 2017.})\cite{fernandes2017internet}. Therefore, it is critical to ensure trustworthy data communication in the IoT.

In the IoT, provenance represents the meta-data that keeps track of the data-generating node (source node) and set of actions performed on data by the intermediate (forwarding) nodes. In RPL network, the UDP packet only contains source and destination fields and thus have no information about the intermediate nodes traversed by the packet. Hence, the malicious nodes can intercept data resulting in false data propagation or false routing information.

We consider the following case studies to illustrate the significance of data trustworthiness in an IoT-enabled paradigm. 

\begin{enumerate}
\item [a)] \textbf{Case Study I} (Nuclear Reactors): System debugging and performance tuning of parameters
\end{enumerate} 
Consider a nuclear reactor where sensors (for instance, temperature, flow, pressure or level) are deployed in order to monitor (heating system, water pressure or water level) and notify sensors' readings (temperature measurements or water level) to a control room (as shown in Fig. \ref{fig:reactor}). The control room is responsible for making critical decisions (to turn on/off any valve or to adjust any values) based on the sensors readings. This working environment may raise the following queries: 
 
    \begin{itemize}
    \item The scientists want to analyze the behavior of the system (throughput) against any particular set of values. How to \emph{tune the parameters} based on digital sensors values?
    \item The scientists want to observe the working of current operations to figure out the problem in system settings. How to \emph{debug the system} by evaluating the sensor values?
    \item A traitor eavesdrops on the sensor data and forged the sensor values (increase or decrease the temperature values) for instance, the destruction caused by Stuxnet worm \cite{kushner2013real} in Iran's nuclear plant. How can the control room \emph{validate the data values} from sensors? 
    \item How to identify that the \emph{disruptions in the system} are either because of internal flaw or any malicious activity?.
    \item How to resolve \emph{intermittent errors} in the system?.
\end{itemize}

\begin{enumerate}
\item [b)] \textbf{Case Study II} (Smart Cities): Availability and Reliability of sensor data
\end{enumerate}
In smart cities, sensors can be deployed in many sectors including monitoring crops for precision agriculture, traffic analysis, environmental (air and water quality) monitoring, energy management, identifying parking spaces, health care, food quality, and many others sensors to control and automate the daily life activities. Setting up such an infrastructure surrounded by sensors requires a careful continuous monitoring of the sensor devices itself. For instance, 
 \begin{itemize}
\item Aging management is required to check for faulty sensors that are giving \emph{abnormal deviations} in results. Similarly, to handle the issues of \emph{intermittent errors} usually more than two sensors are placed in order to get accurate data values and to deal with the case in which one sensor is giving different values than the rest of two sensors.
\item To verify that the sensors in a system are functioning properly requires \emph{debugging} of sensor values.
\item Another major issue is \emph{reliability} of sensor data which may arise due to the cyber threats. \end{itemize}

Under such circumstances, provenance data can help to identify the root cause of disruption in the system. Other supporting examples elucidating the need for provenance are discussed in \cite{suhail2018data}.

\subsection{Challenges and Constraints of a Provenance-Aware IoT System} \label{challenges}
In this subsection, we address the challenges and constraints of a Provenance-Aware IoT system keeping in view the provenance as overhead for IoT devices that are highly constrained in terms of physical size, available memory, CPU power, and battery life. We have also identified requirements and challenges for the integration of secure provenance with IoT in \cite{suhail2016introducing} and \cite{suhail2018data} thoroughly. 

In order to collect meta-data as provenance, it is important to consider the \textit{type of data} to be collected that can help to figure out the data inconsistencies or malicious activities, for example, the ID of participating nodes, time-stamps, sequence number, data flow (destination and source information), etc. The \textit{granularity level of data} being collected further affects the \textit{provenance size}. Provenance size defines that  \enquote{how many extra bytes in a packet are required to be sent along with the data packet as data provenance}. The provenance size ultimately affects \textit{network overhead} and \textit{performance overhead}. Hence, the following two interlinked factors are required to be considered for a provenance-aware IoT system: (i) maintain provenance size to meet the \textit{processing}, \textit{storage} and \textit{energy consumption} constraints of things and (ii) completeness of provenance information to ensure the trustworthiness of data. 

\section{System Model} \label{SystemModel}
In this section, we describe the network model and the data model that we consider for our proposed provenance scheme for RPL networks. We also present the provenance model along with the outline of elemental provenance data components that are utilized in our proposed scheme. Finally, in the adversary model we discuss the anticipated actions a malicious or compromised node can perform to cause disruption in the network. 

\begin{figure}[t!]
\centerline{\includegraphics[width=3.0in]{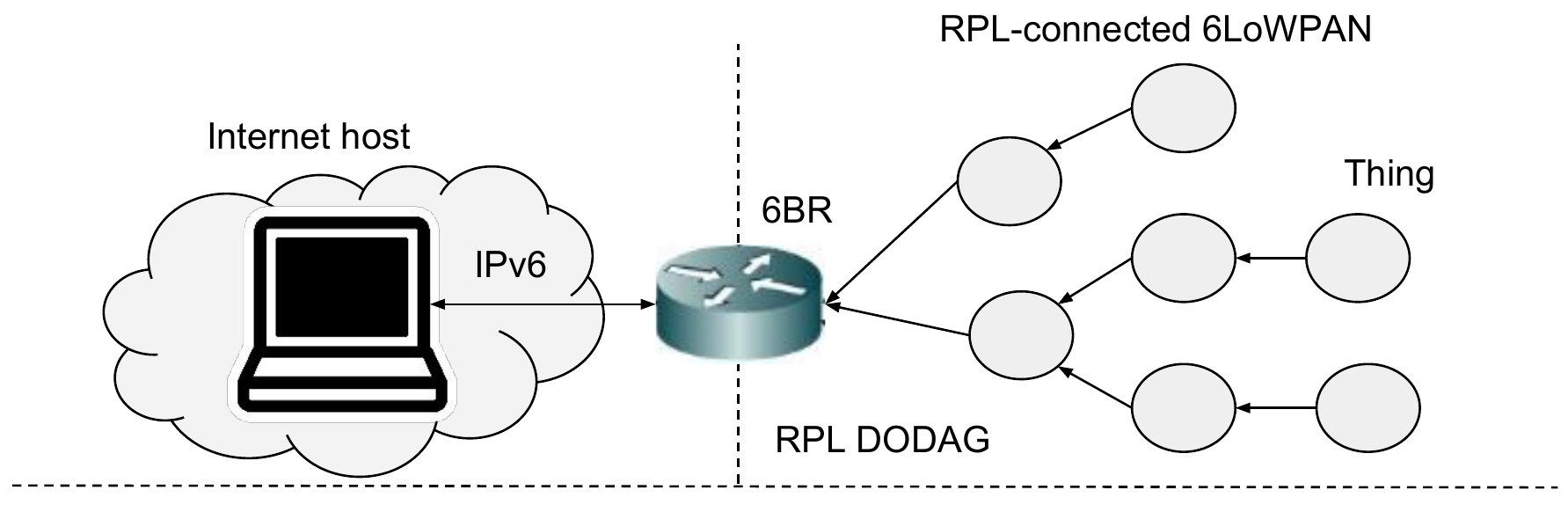}}
\caption{Interconnection of RPL-connected things to the Internet through the Border Router (BR).}
\label{fig:framework}
\end{figure}

\subsection{Network Model}

We consider a RPL network (as shown in Fig. \ref{fig:framework}) where a 6LoWPAN border router (6BR) connects the RPL-connected things in a 6LoWPAN network with the Internet. In RPL-DODAG, a root node (typically a 6BR) performs the task of data collection\footnote{We focus on Multipoint-to-Point (M2P) network scenarios.} from nodes of a sub-DODAG and forwards it to the Internet host that may process, interpret and analyze the sensor data. 

The network is modeled as graph $G (\mathcal{N},R_{info})$ such that:
\[\mathcal{N}= {N_{s}, N_{1} ,\ldots, N_{n}, N_{r},}\]
where $\mathcal{N}$ is the set of nodes consisting of source node $N_{s}$, intermediate nodes (also referred to as forwarding nodes or aggregator nodes) $N_{1} ,\ldots,N_{n}$ and root node $N_{r}$ while $R_{info}$ is the set of routing edges representing the path between the nodes traversed by a data packet i.e., $\langle destination,source \rangle$. 
We consider the following assumptions for our network model:
\begin{itemize}
    
    \item $N_{r}$ has already acquired the knowledge of the packet's path ($\mathcal{P}_{path}$) via DAO messages.
    \item $N_{r}$ has no constraints with respect to energy consumption, storage space, security, and computational/processing capability.
    \item Each non-leaf node maintains a neighbouring table (NT: contains information about parent node, child node(s), and neighbor/sibling nodes (see Fig. \ref{fig:DODAG}) and a routing table (RT: contains information about node child node(s) and their respective child nodes (see Fig. \ref{fig:DODAG}).
    \item Routing paths may change over time due to node failure, mobility, link quality degradation, congestion, resource optimization etc. Hence, our model adapts itself to both static and dynamic scenarios of a sensor network.
\end{itemize} 
\subsection{Data Model}
We assume that some of the sensor nodes referred to as \emph{source nodes} ($N_s$) in a network generate data packets ($d_{p}$) periodically. $N_s$ forwards $d_{p}$ to another node referred to as \emph{forwarding node} ($N_f$) that appears next on the path. The packet path ($\mathcal{P}_{path}$) that is the path traversed by $d_{p}$ can be represented as: 
    \[\mathcal{P}_{path}= {N_{s}\rightarrow N_{pp1}\rightarrow,\ldots,\rightarrow N_{ppn} \rightarrow N_{r}},\]
    i.e., the source child node (data-generating node) forwards the packet to its preferred parent $pp$ until it arrives at the root node.
For instance, in Fig. \ref{fig:DODAG} node 7, 8, 9 and 10 are data source nodes, node 2, 3, 4 and 5 are forwarding nodes while node 6 aggregates data received simultaneously from multiple node and acts as an \emph{aggregator node}. We assume aggregator node similar to a forwarding node as it is maintaining routing information for each of the source node or forwarding node.

\begin{figure}[t!]
\begin{tikzpicture}[node distance=2.2cm, every node/.style={fill=white, font=\sffamily}, align=center]
       \node (start)             [activityStarts]              {Data};
      \node (onData)     [process, right of=start]          {$Seq\#$};
      \node (onSeq)      [process, right of=onData]   {$P_D$};
      \node (onData)     [process, below of=onSeq]          {$R_{info}$};
      
    \draw[->]             (onSeq) -- (onData);
    \node (onData2)     [process, below of=onData]          {$\langle destination\ node\ ID, source\ node\ ID \rangle$};
      
    \draw[->]             (onData) -- (onData2);
\end{tikzpicture}
\caption{Payload entries: data, sequence number, and provenance data i.e., routing information as $\langle destination\ node\ ID, source\ node\ ID \rangle$ pair. }
\label{fig:payload}
\end{figure}
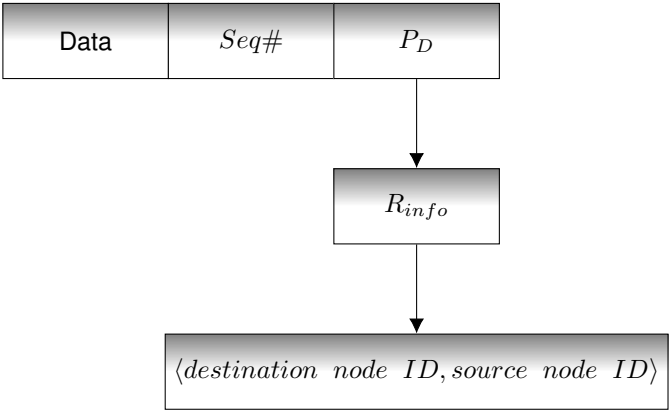

The main components of a $d_{p}$ are: (i) a unique sequence number $S$, (ii) data (also known as payload) $d$, (iii) packet path $\mathcal{P}_{path}$ refer to as provenance information or routing information ($R_{info}$) (shown in Fig. \ref{fig:payload}), and (iv) Message digest (MD) of $R_{info}$.
We have used SHA-2 for computing the MD. The MD included in the data packet is not used in provenance encoding or decoding process. Hence, it is not a part of the provenance and is used only to ensure the provenance’s integrity.

\subsection{Provenance Model}
We consider a provenance-aware data model that keeps track of:
\begin{enumerate}
\item [a)] Set of participating nodes ($N_{\omega}$) where $N_{\omega}$ $\subset$ $\mathcal{N}$ having $\langle Parent,Child \rangle$ relationship.
\item [b)] Data packet $d_{p}$ along with routing information\footnote{$R_{info}$ is represented as $\langle Parent,Child \rangle$ relationship.} ($R_{info}$) encapsulated at each forwarding node $N_{f}$ traversed by $d_{p}$. 

\item [c)] Sequence number $S_{i}$ (extracted from $d$) against the routing entry\footnote{$R_{e}$ is represented as next hop child node $\langle Child \rangle$.} ($R_{e}$) in routing table (RT) at each forwarding node $N_{f}$. 
\end{enumerate}

Hence, provenance data $P_{D}$ can be represented as: 
\begin{equation}
 R_{info}\ @\ d_{p}\ ||\ S_{i}\ @\ RT.
\end{equation}

Thus, $R_{info}$ represents the routing information in terms of $\langle Parent,Child \rangle$ relationship and $S_{i}$ represents $i^{th}$ sequence number of $i^{th}$ data packet $d_{i}$ maintained at RT of $N_{f}$. It is important to note that $||$ operator means that both routing information and sequence numbers are stored in the data packet and the routing table respectively and combination of both can be used to trace packet path, packet drop or faulty node during provenance decoding. $@$ operator means that $R_{info}$ is embedded in $d_p$ while $S_{i}$ is maintained at RT.

The provenance data ($P_{D}$), more specifically, in terms of node-level and system level provenance can be represented as: 
\begin{subequations}\label{2}
\begin{equation}\label{2a}
RT \leftarrow S_{i}, \qquad \textit{node-level provenance,}
\end{equation}
\begin{equation}\label{2b}
d_{p} \leftarrow (R_{info}=\langle N_p, N_c \rangle), \qquad \textit{system-level provenance.} \qquad
\end{equation}
\end{subequations}

Thus, each $N_{f}$ in $\mathcal{P}_{path}$ keeps on updating $R_{info}$ and embeds it in the payload of $d_{p}$ while $S_{i}$ is embedded against
$R_{e}$ in RT.

\subsection{Adversary Model}
During the process of network operations (i.e., topology formation and packet exchanging), only $N_r$ is considered to be a trusted entity. An adversary in the network may compromise other benign nodes. The malicious node ($N_m \in N_{\omega}$) can drop, inject, alter, replay or forge data packets that ultimately imperils the trustworthiness of data. Here, we have particularly consider $N_m$ impersonating as $N_f$. $N_r$ may not be able to distinguish between benign or compromised nodes. Hence, $N_r$ needs to verify the reliable traversal of data through $P_{D}$. In our proposed scheme, we do not emphasize on the confidentiality of provenance data. However, confidentiality can be achieved by encrypting the provenance data by using the secret keys of the respective nodes. Our proposed scheme aim to achieve the following
security properties:
\begin{enumerate}[a)]
    \item Any malicious or faulty nodes are detectable.
    \item Any malicious node can not remove provenance (node-level or system-level) from the data packet without being detected.
    \item A malicious node cannot inject counterfeit information in the provenance record or remove any benign node from the provenance record. 
\end{enumerate}
\section{Provenance Scheme} \label{scheme}
Due to the low power and lossy nature of RPL networks, the data packets are subjected to packet loss because of network disruptions (congestion, link failure) or spurious attacks by malicious or compromised nodes. To identify problematic or faulty node, the first step is identifying the source node followed by identifying the forwarding nodes. However, it requires to adopt some preliminary steps to be taken at the time of \textit{data generation} and \textit{data collection}. In Provenance-enabled Packet Path Tracing (PPPT) scheme, we have introduced a path correlation provenance scheme that generates provenance efficiently in terms of 
i) the routing entry tagged with the sequence number of a data packet,  
ii) routing information based on packet traversal. 

In this section, we provide details of provenance encoding and decoding using flowcharts, algorithms and a working example.  
\begin{table}[t!]
\begin{center}
\caption{Notations}
\label{tab:symbols}
 \begin{tabular}{c c } 
 \hline
 \hline
\textbf{Symbol} & \textbf{Description} \\ [0.5ex] 
 \hline
 $P_{size}$ & Provenance size  \\ 
 $P_D$ & Provenance data  \\ 
 $d_p$ & Data packet  \\ 
 $\mathcal{P}_{path}$ & Packet path \\
 $R_{info}$ & Routing information  \\
 $R_e$ & Routing entry \\
 $pp$ & Preferred parent\\
 $N_s$, $N_f$, $N_r$  & Source node, Forwarding node, Root node  \\
 $N_p$, $N_c$ &  Parent node, Child node \\
\hline
\hline
\end{tabular}
\end{center}
\end{table}

\begin{figure*}[t!]
\centering
\begin{subfigure}[b]{0.48\textwidth}
\centering
\includegraphics[width=2.5in]{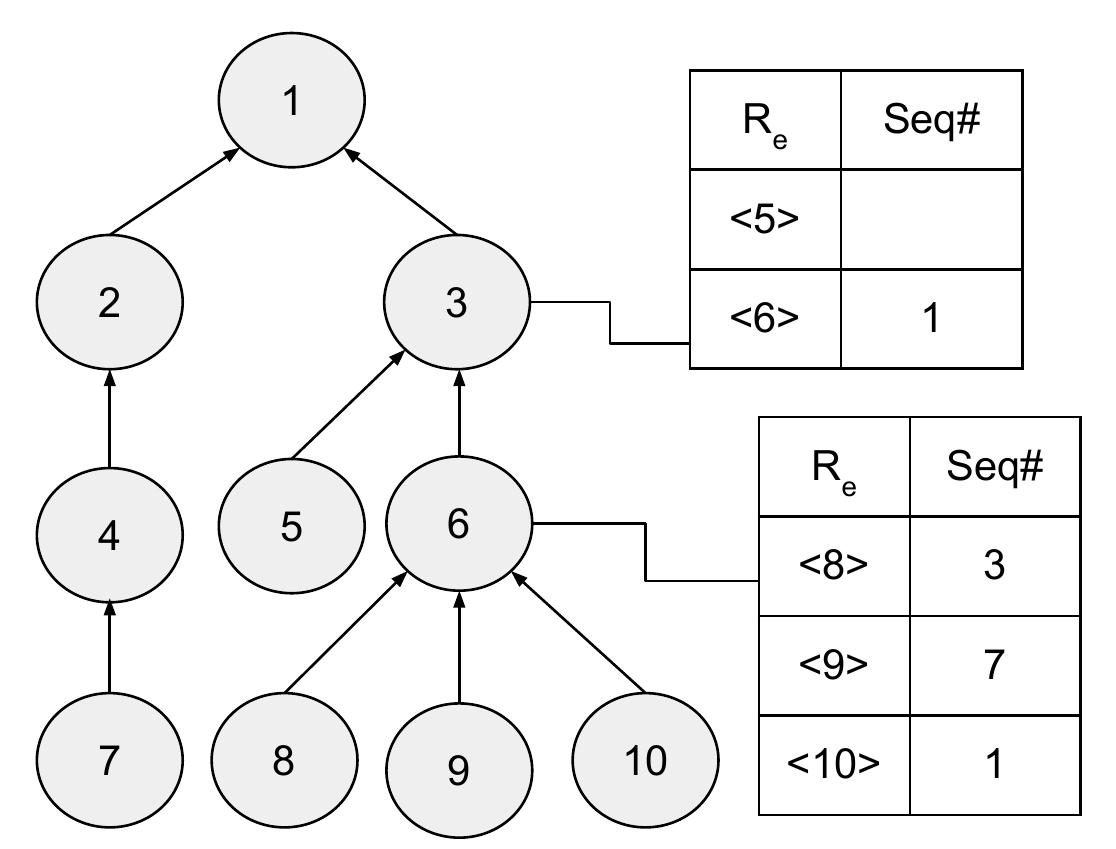}
\caption{$S$ against $R_e$}
\label{fig:RT}
\end{subfigure}%
\hspace*{\fill}
\begin{subfigure}[b]{0.48\textwidth}
\centering
\includegraphics[width=2.5in]{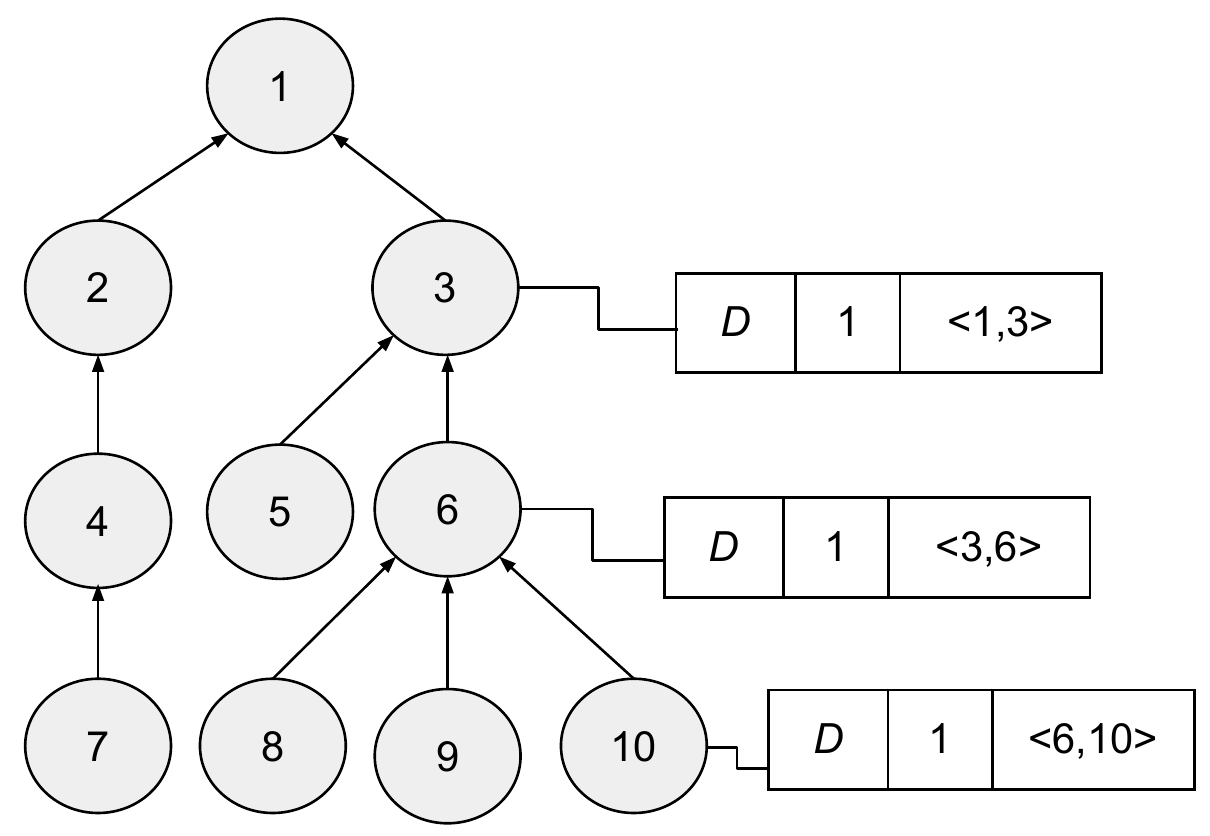}
\caption{$R_{info}$ in $d_p$}
 \label{fig:Rinfo}
\end{subfigure}
\centering
\caption{ Provenance encoding: (a) embedding sequence number against routing entry in the RT of preferred parent. (b) enclosing routing information in data packet at each forwarding node. }
\label{fig:packet_routing}
\end{figure*}

\subsection{PPPT: Provenance Encoding}
In RPL network topology, each parent node $N_{p}$ maintains a route entry $R_{e}$ of its child node(s) $N_{c}$ in a routing table (RT). RT is a table at each $N_{f}$ (when downward routes are enabled) that keeps track of the routes originating from child nodes in the form of $\langle Parent,Child \rangle$. During the process of establishing DODAG topology (as discussed in Section \ref{rpl-topology}), RT is populated with routing entry based on next hop node as $\langle Child \rangle$. To track the detection of lost packets and the malicious nodes responsible for packet loss in the network we have introduced another field \textit{sequence number $S_{i}$} that is added against each $R_{e}$ in RT of node $N_{i}$. In addition to this, $R_{info}$ is embedded along with the payload (as shown in Fig. \ref{fig:packet_routing}) in terms of $D_{id},S_{id}$ ($\langle destination\ node\ ID, source\ node\ ID \rangle$) at each $N_{f}$. Unlike existing research works that employs separate transmission channels for data and provenance \cite{simmhan2005survey}, we only require a single channel for both. 

\begin{figure}[t!]
\centerline{\includegraphics[width=3.5in]{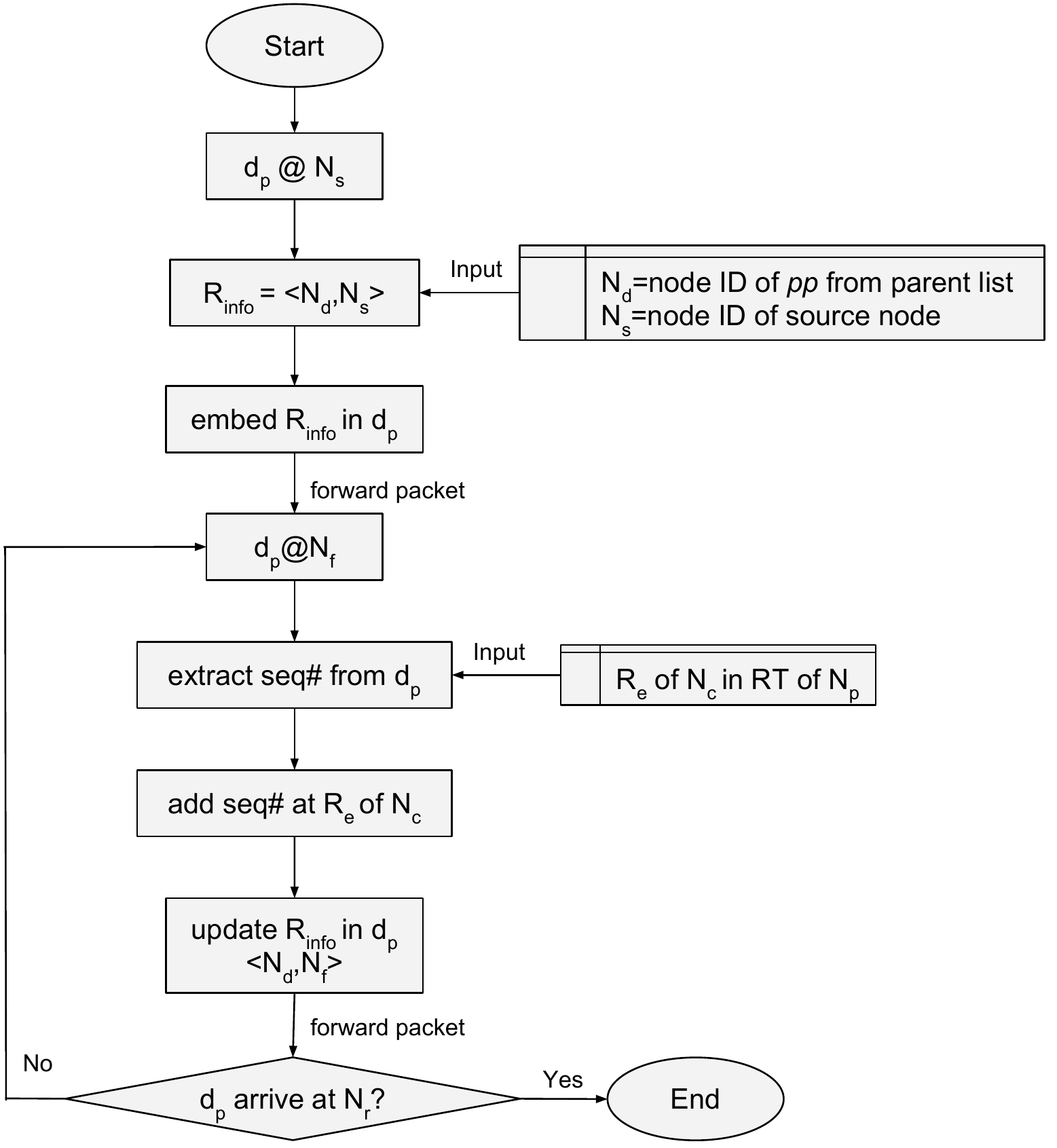}}
\caption{Provenance encoding workflow at the participating nodes in the network.}
\label{fig:flowchart_embed}
\end{figure}

\begin{algorithm}[t!]
\caption{Provenance Encoding}
\label{algo:encoding}
\begin{algorithmic}
\Require $d_{p}$
\Ensure $R_{info}$@$d_{p}$ , $S_{i}$@RT
    \For{\texttt{each source node $N_s$}}
        \State \texttt{$N_s$ elects $pp$ from its parent\_list}
        \State \texttt{$N_d$=$N_{id}$ of $pp$}
        \Comment{$N_d$ refers to as next hop $pp$.}
         \State \texttt{$N_s$=$N_{id}$ of $N_s$ }
         \State \texttt{$R_{info}$ $\leftarrow$ $\langle N_{d},N_{s} \rangle$}
          \Comment{insert $N_{d}$ and $N_{s}$ into $R_{info}$. }
        \State \texttt{$d_{p}$ $\leftarrow$ $R_{info}$ } 
         \Comment{embed $R_{info}$ in $d_{p}$ payload.}
        \State \texttt{forward $d_{p}$ to $pp$} 
    \EndFor
    \For{\texttt{each forwarding node $N_{f}$}}
        \State \texttt{extract $S_{i}$ from payload of $d_{p}$}
        \State \texttt{go to RT of $N_{f}$}
        \State \texttt{look out for $R_e$ of $\langle N_c \rangle$ }
        \Comment{$R_e$ refers to next hop child node in RT.}
        \State \texttt{$R_{e}$@RT of $N_{f}$ $\leftarrow$ add\_seq($N_{c}$)}
        \State \texttt{$R_{info}$ $\leftarrow$ $\langle N_{d},N_{f} \rangle$}
        \Comment{$N_f$ refers to as immediate $N_c$.}
         \State \texttt{update $R_{info}$ in $d_{p}$ payload}
   \EndFor
\end{algorithmic}
\end{algorithm}
To illustrate the working of provenance encoding let us consider the topology shown in Fig. \ref{fig:packet_routing}. Suppose $N_{10}$ is a data-generating node. To send a packet, $N_{10}$ embeds $R_{info}$ as $\langle 6,10 \rangle$ in $d_p$ and forwards the packet to its preferred parent ($pp$) $N_{6}$. Upon packet reception, 
(i) $N_{6}$ extracts the sequence number 1 from payload and inserts in it's RT against $R_{e}$ $\langle 10 \rangle$, 
(ii) it updates $R_{info}$ as $\langle 3,6 \rangle$, 
(iii) it forwards the packet to its $pp$ i.e., $N_{3}$. $N_{3}$ follows the similar steps by adding the sequence number against $R_{e}$ $\langle 6 \rangle$ (as shown in Fig. \ref{fig:RT}) and updating $R_{info}$ (as shown in Fig. \ref{fig:Rinfo}) until the packet arrives at $N_{1}$. Thus, $P_{D}$ is maintained at RT in terms of $S_i$ against $R_{e}$ at respective $N_{f}$ in RT while $R_{info}$ is maintained at $d_{p}$. Fig. \ref{fig:flowchart_embed} shows provenance encoding workflow at the participating nodes. 

\begin{figure}[t!]
\centerline{\includegraphics[width=3.2in]{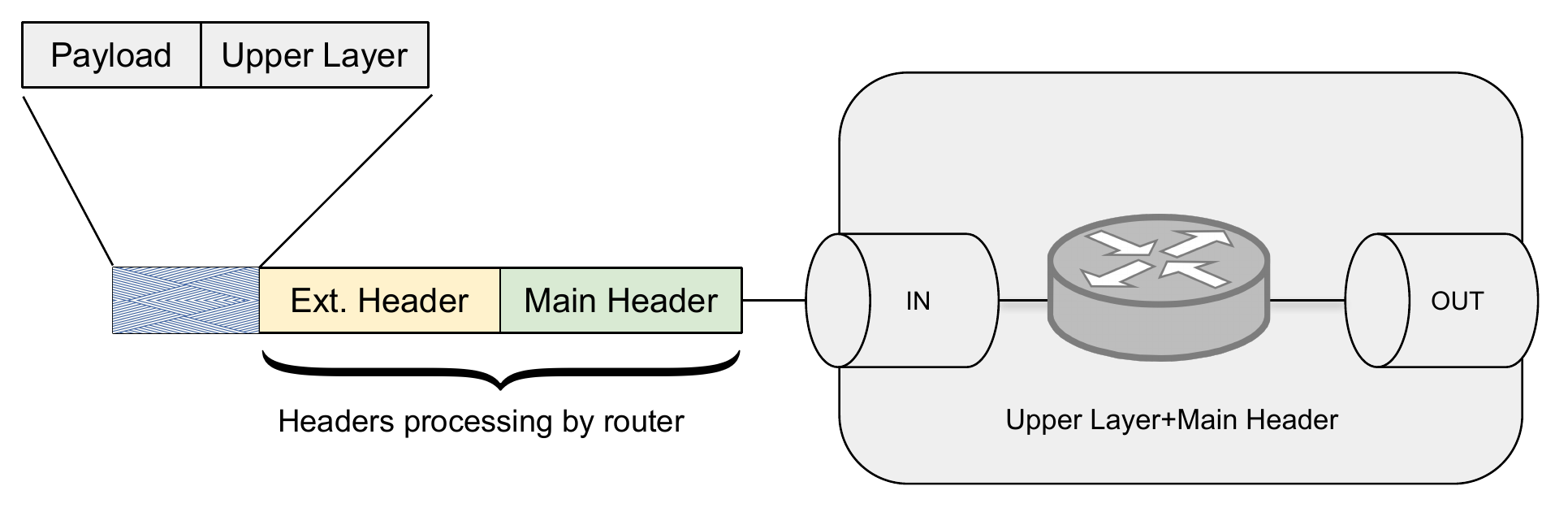}}
\caption{Headers processing at sink node.}
\label{fig:ipv6}
\end{figure}

To provide a quick \textit{provenance assessment} check at $N_r$, we set a flag in the IPv6 extension header (EH) by utilizing the available reserved bits. The bit is set at the time of provenance embedding and is recovered upon packet reception at the sink. The primary rationale of setting this bit is to inform $N_r$ about the presence of $P_D$ in the payload. As $N_r$ processed the main IPv6 header and EH first so it can know about the presence of $P_D$ through the set bit. Moreover, if any malicious node has dropped $P_D$ from the payload then $N_r$ can know about modification in the payload and can simply discard the packet without further processing. Thus, the set bit in EH enables efficient provenance decoding and data trust assessment. Fig. \ref{fig:ipv6} shows the processing order of headers by $N_r$ or 6BR. 

\begin{figure*}[ht!]
\centerline{\includegraphics[width=3.5in]{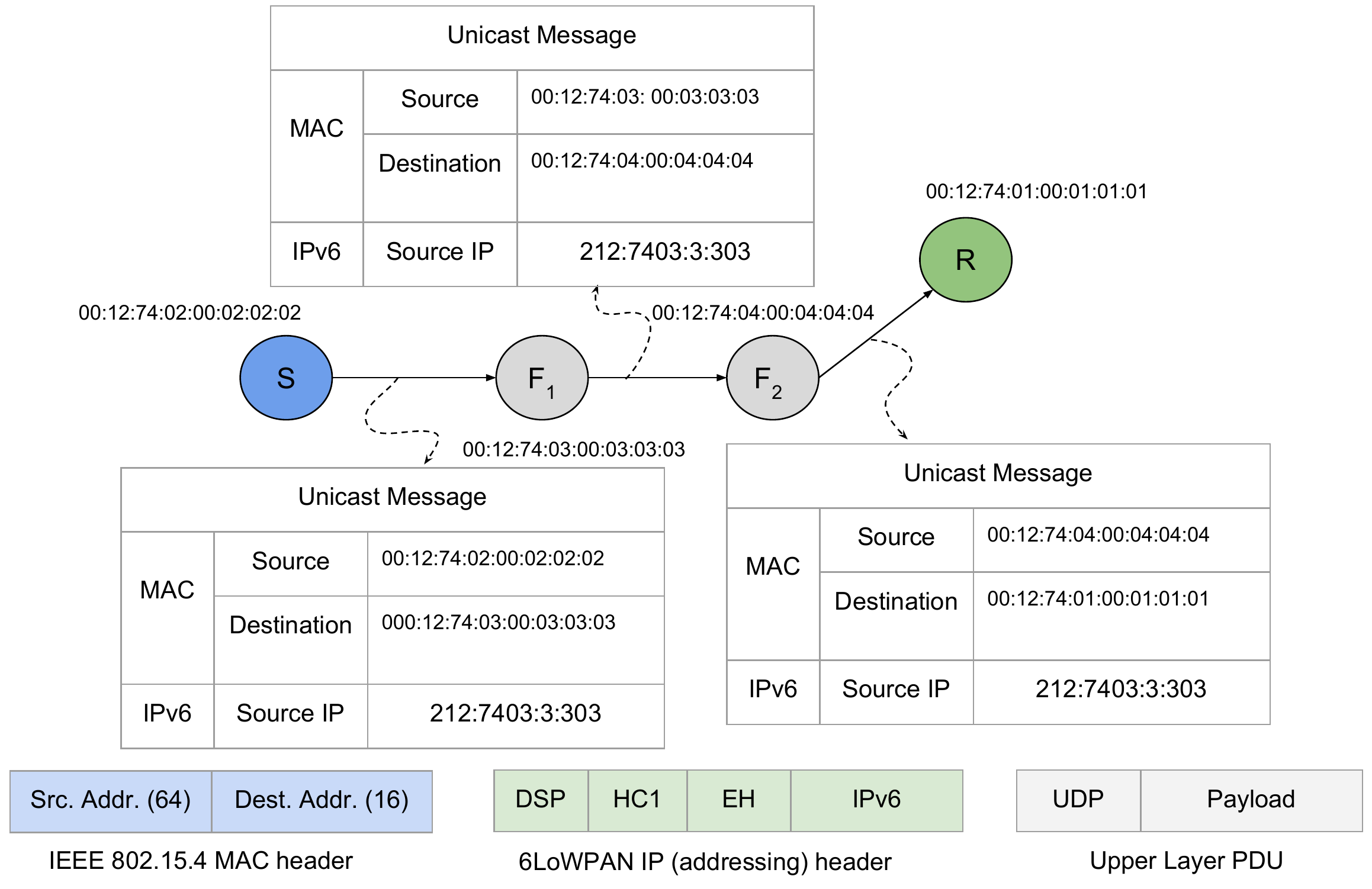}}
\caption{Layer-2 packet forwarding scenario from source to destination through intermediate nodes in 6LoWPAN network.}
\label{fig:ipv6_detail}
\end{figure*}

Fig. \ref{fig:ipv6_detail} provides a detailed overview of the layer-2 packet forwarding scenario from the source node $S$ to the destination node $R$ via intermediate nodes $F_1$ and $F_2$. The forwarded unicast message includes MAC header, 6LoWPAN header and Upper layer protocol data units (PDUs). During each traversal, the intermediate node will look up in its MAC address table to forward the unicast payload message to the neighboring/preferred parent node until the packet arrives at the root node.

Another way is to embed the node ID of each participating node in the packet traversal called as Provenance as ID (PID) \cite{suhail2018data}. However, it will increase the provenance size exponentially at each hop count. We compare PPPT with PID in Section \ref{results}. In the case of PID, $R_{info}$ takes the form $\langle 10,6,3,1 \rangle$ after arriving at $N_r$. 

$R_{info}$ represents \textbf{\textit{system-level\ provenance}} as it gives a sheer picture of the participating node(s) in the network. It is lossless because it is recording the complete ordered packet traversal information. The root node extracts $R_{info}$ from $d_{p}$ for performing provenance decoding (discussed in Section \ref{Pverify}). 
On the other hand, maintaining sequence number against $R_{e}$ in RT acts as \textbf{\textit{node-level provenance}}. The main goal of maintaining such information at each node is to point out the faulty node(s) or compromised node(s) that are causing the packet drop attack. For instance, congestion caused in the case of event regions as discussed by \cite{khan2016sink} can be resolved by employing provenance information. Similarly, a compromised node can be identified based on packet count or packet delivery ratio (PDR) as it is causing abnormal packet drop \cite{suhail2018detection}. 
Considering the memory issues of resource-constrained nodes, we have associated sequence numbers with an interval $I$. At the end of each round interval, the sequence numbers are reset at the RT of non-root nodes, thus making room for newer data packets. Since $N_r$ does not have any constraint related to storage and processing, therefore, we don't need to reset the sequence number entries in the RT of the root node. Moreover, the complete provenance information can be checked from the root node whenever required for system debugging etc. 

\begin{figure}[t!]
\centerline{\includegraphics[width=3.5in]{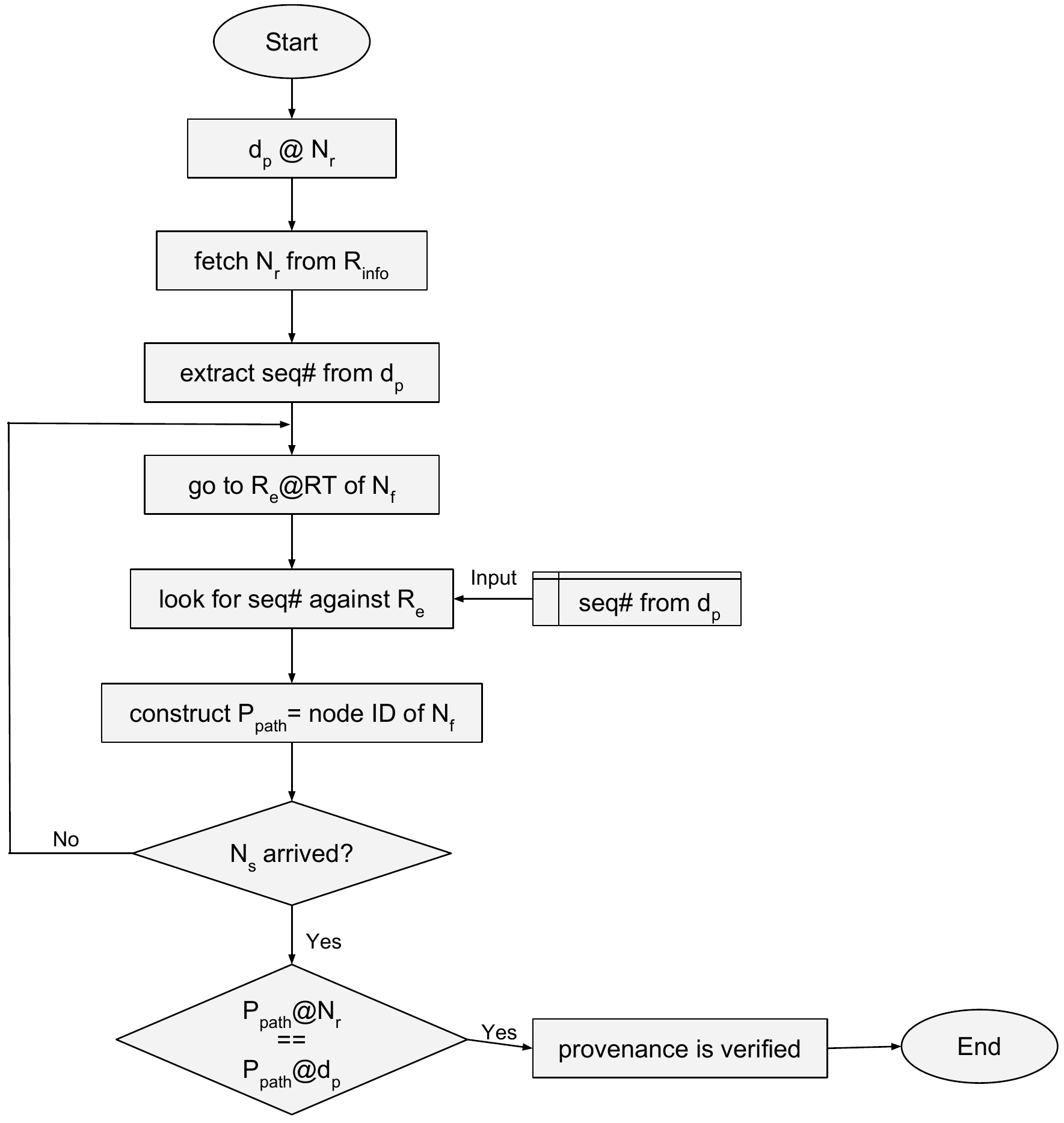}}
\caption{Provenance processing workflow at the root node upon receiving a data packet.}
\label{fig:flowchart_decode}
\end{figure}
It is important to note that the root node is maintaining the information about all routing entries through which packets have been traversed. For example, in the case of node failure, root node will update its table entries accordingly by broadcasting DIO messages and thus enabling nodes to select their preferred parent. In other words, the root node is aware of the source nodes that generate the data, forwarding nodes that forward the packets and the packet path traversed by the packet. The knowledge acquired by the root node during topology formation (or network establishment) is used during the provenance verification process at the root node. 
\subsection{PPPT: Provenance Decoding} \label{Pverify}
The root node upon packet reception performs the provenance verification by extracting sequence number and routing information from the data packet by exploiting both node-level and system-level provenance. Firstly, it extracts both value i.e., $R_{info}$ and $S_i$ from the payload of $d_p$. Secondly, based on $S_i$, $N_r$ go to the RT of $N_f$ and look out for $R_e$ against $S_i$. $N_r$ continues the process until it traverses the packet path and accumulates it in $\mathcal{P}_{path}$. $N_{r}$ is able to trace the packet traversal (as $\mathcal{P}_{path}@N_r$) through its RT that contains an exhaustive view of the entire network topology. Thirdly, after accumulating information in $\mathcal{P}_{path}@d_p$, $N_r$ compares both $\mathcal{P}_{path}@N_r$ and $\mathcal{P}_{path}@d_p$. The provenance is verified if both values are equal, otherwise $N_r$ discards the packet. 
The decoding steps are discussed in algo. \ref{algo:decoding}. The workflow shown in Fig. \ref{fig:flowchart_decode} summarizes the provenance decoding process.

To explain the decoding process we again make reference to Fig. \ref{fig:RT}. 
$N_1$ takes $\langle N_{3} \rangle$ from $R_{info}$ pair $\langle N_{1},N_{3} \rangle$. It then goes to the RT of $\langle N_{3} \rangle$ and look out for the sequence number 1 maintained against $R_e$ which is $\langle N_{6}\rangle$. To further carry on the packet traversal, $N_r$ goes to the RT of $\langle N_{6} \rangle$. Here again it look out for sequence number 1 against $R_e$ which is $\langle N_{6}\rangle$. Finally, it reaches the source generating node i.e., $\langle N_{10}\rangle$. After completion, it stores the result in $\mathcal{P}_{path}@d_p$ which is then compared with $\mathcal{P}_{path}@N_r$ for provenance verification.  
\begin{algorithm}[t!]
\caption{Provenance Decoding}
\label{algo:decoding}
\begin{algorithmic} 
\Require $S_{i}$, $R_{info}$, $\mathcal{P}_{path}@N_r$
     \For{\texttt{each $d_{p}$} @ $N_{r}$} 
        \State \texttt{fetch $N_f$ from $R_{info}$}\Comment{system\_level provenance.}
        \State \texttt{extract $S_i$ from $d_{p}$}\Comment{node\_level provenance.}
             \For{\texttt{each $S_i$}}
             \State \texttt{go to $R_e$@RT of $N_f$} \Comment{traverse $\mathcal{P}_{path}$. }
             \State \texttt{look for $S_i$ against $R_e$ of $N_c$ in RT}
            \State{$\mathcal{P}_{path}$@$d_p$ $\leftarrow$ $N_{id}$ of $N_f$}
           \Comment{construct packet path. }
            \EndFor
           
    \EndFor
        
        \If{$\mathcal{P}_{path}@N_r$==$\mathcal{P}_{path}@d_p$} 
        \State {return true} \Comment{provenance is verified.}
       \Else
        \State {return false}
        \EndIf

\end{algorithmic}
\end{algorithm}
\section{Security Analysis} \label{claims}
In this section, we discuss the security claims and their justification to evaluate the performance of our proposed provenance scheme.

\textit{Claim 1: Faulty nodes are detectable}.

\textit{Justification}: As a part of system-level provenance sequence number is embedded in the payload. Also as a part of node-level provenance, the sequence number is maintained at the parent node RT against the respective child nodes as routing entries. The sequence number can be tracked by the root node at each forwarding node for detecting a faulty or a malicious node. Therefore, any malicious node cannot selectively drop data packets without being identified.

\textit{Claim 2: A malicious node can not remove provenance (node-level or system-level) from the data packet without being detected}.

\textit{Justification}: Any malicious node may attempt to remove the provenance data from the data packet. However, the flag in the IPv6 EH is set showing the presence of provenance in the data packet. Hence, upon the packet reception at the sink node, the flag bit can provide an idea to the sink node about the absence of provenance data in the data packet. The sink node can simply discard the packet without further processing.  

\textit{Claim 3: A malicious node cannot inject counterfeit information in the provenance record or remove any benign node from the provenance record}.

\textit{Justification}:
Since the sequence number is being inserted in the routing table against the routing entries hence, even if the adversary forges the routing information the routing entries hold the information about the sequence numbers along with routing entries. 

\section{Simulation}\label{simulation}
In this section, we have evaluated the performance of our proposed provenance-enabled packet path tracing scheme (PPPT) through simulation for linear topology. We analyze the performance of PPPT with respect to provenance size, provenance generation time, memory consumption (ROM and ROM overhead), robustness, and energy consumption. 
\begin{table*}[t!]
\begin{center}
\caption{Network parameters used in simulation analysis} \label{tab:parameters}
  \begin{tabular}{|c|c| } 
 \hline
 \rowcolor{lightgray!30}
\textbf{Parameter} & \textbf{Value} \\ [0.5ex] 
 \hline
 Network layer & RPL  \\ 
 MAC layer & 802.15.4  \\ 
 Simulation time & 600 s  \\ 
 Radio medium & Unit Disk Graph Medium(UDGM):Distance Loss\tablefootnote{UDGM:Distance Loss models the transmission range as a circle in which only the nodes inside the circle receive packets while considering interference.}\\
 Topology & Linear\\
 Number of forwarding nodes & varying from 1 to 7\\
 Packet size (excluding header) & 200 bytes\\
 Data rate & varying from 1 packet/10 sec to 1 packet/40 sec\\
\hline
\end{tabular}
\end{center}
\end{table*}
\subsection{Simulation Setup}
We run our experiments in Contiki-based network simulator Cooja \cite{eriksson2009cooja}.
Contiki \cite{dunkels2004contiki} has a well tested implementation of RPL (ContikiRPL). For our simulations, we use Tmote Sky \cite{tmote} running ContikiMAC as things in 6LoWPAN networks. ContikiMAC is a default Contiki RDC (Radio Duty Cycling) protocol that allows nodes to keep their radio off for most of the time (> 99\%) and switch it on only when needed to conserve energy while being able to relay multi-hop messages. Tmote Sky uses CC2420 IEEE802.15.4 transceiver and has 48KB of flash and 10KB of RAM. The parameters associated with the simulation are shown in Table \ref{tab:parameters}.

\subsection{Performance Metrics} 
We analyze the performance of our proposed provenance scheme using the following performance metrics:
\subsubsection{Provenance Size}
Any additional information or meta-data used to identify the data source and packet path constitutes provenance data ($P_{D}$). The number of extra bytes in the data packet (as $P_{D}$) is referred to as provenance size ($P_{size}$) and can be defined as:
\begin{equation} \label{eq:psize}
\begin{aligned}
P_{size}\ (bytes){}=Payload-P_D,
\end{aligned}
\end{equation}
where as $P_D$ can be represented as:
\begin{equation} \label{eq:pdata}
\begin{aligned}
P_{D}\ (bytes){}=R_{info}\ @\ N_i.
\end{aligned}
\end{equation}
$P_{size}$ is considered to be a significant factor in the evaluation of PPPT as it justifies the performance overhead of the entire network.  \\

 In the case of PPPT, we only need to maintain tracing information at each node as:
\[\langle next\_destination\_node, current\_source\_node \rangle, \]  
Hence, $P_{size}$ remains constant throughout the packet traversal and is irrespective of the number of hops count (i.e., number of nodes to be traversed by the data packet to reach the destination/sink node) or the number of nodes in the network. However, in the case of PID, it maintains information about each traversed node as:
\[\langle source\_node, forwarding\_node(s), destination\_node \rangle. \]
Hence, $P_{size}$ increases with the increase in the hop count. 

\subsubsection{Energy Consumption}
Due to the resource-constrained nature of IoT network, we have to consider network-level energy consumption and node-level power consumption incurred due to the inclusion of provenance information. In order to estimate power, \textit{PowerTrace} and \textit{Energest} tools included in Contiki OS has been used. The time each mote spends in the states including transmitting time ($T_{x}$) (i.e., the radio is transmitting with the
MCU on), receiving time ($R_{x}$) (i.e., the radio is receiving with the MCU on), low power mode (LPM) (i.e., the MCU is idle and the radio is off), and processing/CPU time (the MCU is on and the radio is off) can be determined through these tools. 

Based on the nominal values (the typical operating conditions
of the Tmote Sky) available in \cite{tmote}, the network-wide energy usage for 600 sec by all the nodes can be calculated as:
\begin{equation} \label{eq:energy}
\begin{aligned}
Energy\ (mJ){}=(Tx * 19.5 mA+ Rx * 21.8 mA\\
+CPU * 1.8 mA+ LPM * 0.0545)*3 V \\
/32768*8.
\end{aligned}
\end{equation}

From the network-wide energy usage, we calculate the average power as:
\begin{equation} \label{eq:power}
\begin{aligned}
Power\ (mW)=\frac{Energy\ (mJ)}{Time\ (s)}.
\end{aligned}
\end{equation}

To compute per-node average power consumption for the node's active time period, power is divided by the total number of nodes. We have measured energy consumption by using both radio duty cycling and non-duty cycling. We have used \textit{contikimac\_driver} for duty cycling and \textit{nullrdc\_driver} for non-duty cycling.

\subsubsection{Packet Drop Detection Rate (PDDR)}
The packets are subjected to loss due to various anomalies including node failure, mobility, or link failure during transmission. In addition to the malicious activity, packet loss may also occur even in the case of benign network situation. 
By computing the ratio between the number of packets detected as dropped to the number of packets that are actually dropped, we can evaluate the robustness of PPPT. Following this correlation, if $l_p$ represents detected dropped packet and $t_p$ represents actual dropped packet then PDDR can be computed as:

\begin{equation} \label{eq:PDDR}
\begin{aligned}
PDDR{}=\frac {l_p}{t_p}.
\end{aligned}
\end{equation}
\begin{figure*}[ht!]
\begin{center}
 \begin{subfigure}[b]{0.48\textwidth}
\includegraphics[width=2.5in,height=5.2cm]{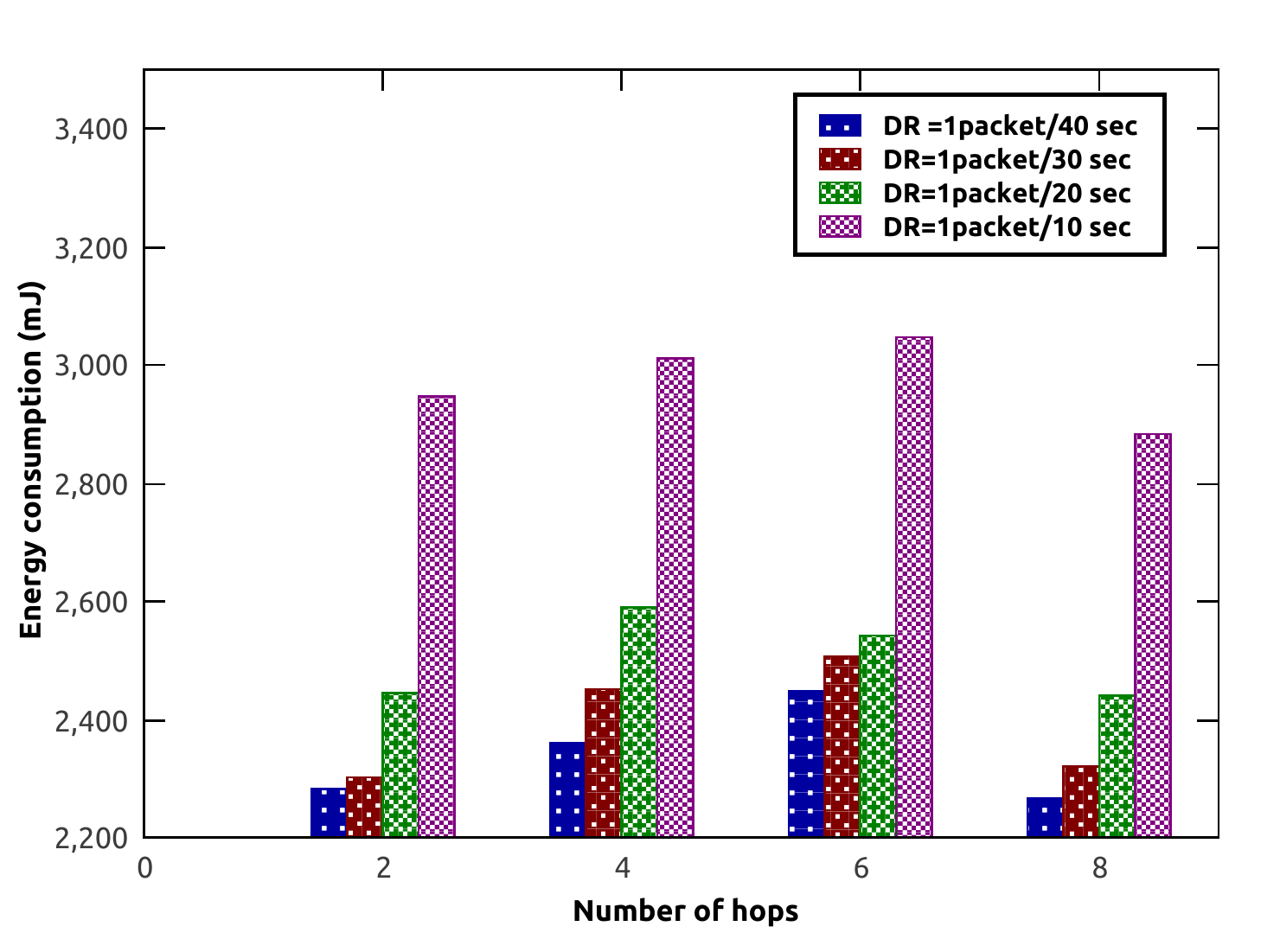}
\caption{Energy usage for the entire network with duty-cycling.}
\label{graph:E_With_DutyCycling}
\end{subfigure}%
\hspace*{\fill}
 \begin{subfigure}[b]{0.48\textwidth}
\includegraphics[width=2.5in,height=5cm]{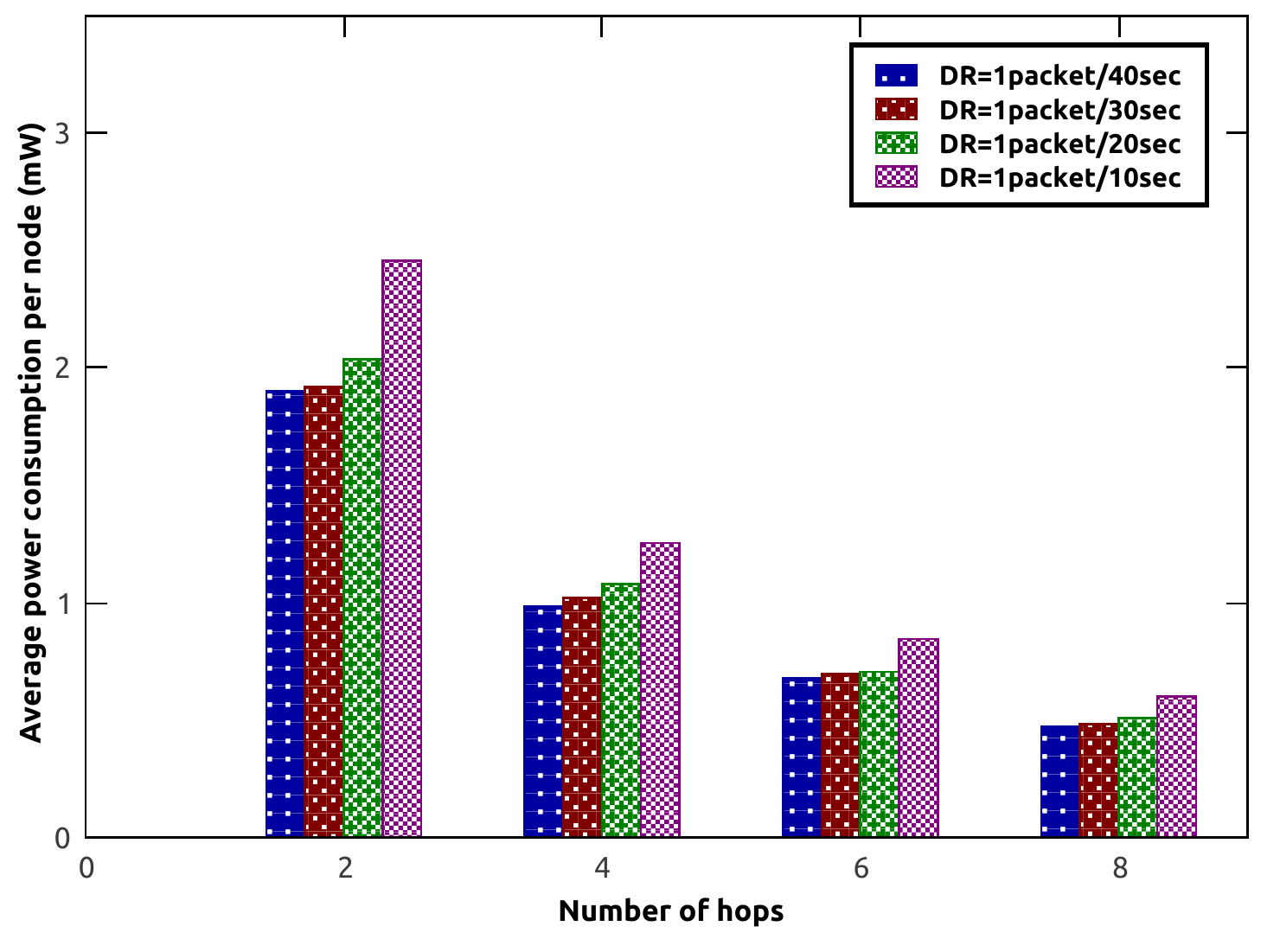}
\caption{Average power usage per node with duty-cycling.}
 \label{graph:P_With_DutyCycling}
\end{subfigure}
\caption{ (a) Network-level energy consumption with respect to varying data rates at different hop counts. (b) Node-level average power consumption with respect to varying data rates at different hop counts. }
\label{graph:consumption}
\end{center}
\end{figure*}
\subsubsection{Average Provenance Generation Time}
Provenance generation time can be defined as the time (including the other network processing delays) required by the forwarding node to embed provenance and forward the data packet to the next neighboring or preferred parent node.
\begin{equation} \label{eq:pgt}
\begin{aligned}
PGT(min)=t_{n}-t_{n-1},
\end{aligned}
\end{equation}
where $t_n$ and $t_{n-1}$ represents the time taken by $n^{th}$ node and $n-1^{th}$ node respectively. 
The average provenance generation time can be computed as follows:
\begin{equation} \label{eq:avgpgt}
\begin{aligned}
Average\ PGT(min)=\frac{\sum\limits_{i=1}^m PGT_{n_i}}{m},
\end{aligned}
\end{equation}
where $PGT_{n_i}$ represents the time consumed by a node $n_i$ and $m$ represents the number of packets being forwarded.
\begin{table}
\begin{center}
\caption{Additional RAM and ROM required by PPPT.} \label{tab:resource}
  \begin{tabular}{|c|c|c| } 
 \hline
 \rowcolor{lightgray!30}
\textbf{Overhead} & \textbf{RAM (Bytes)} & \textbf{ROM (Bytes)} \\ [0.5ex] 
 \hline
 Additional storage & 504 & 3874  \\ 
\hline
\end{tabular}
\end{center}
\end{table}

\subsection{RAM and ROM Usage}
Since IoT devices have limited ROM and RAM resources, therefore it is important to ensure that the additional overhead in terms of provenance is optimized for these environments. In order to quantify ROM and RAM usage by our proposed scheme, we have used the \textit{msp430-size} tool. Table \ref{tab:resource} shows the RAM and ROM overhead of provenance-based RPL scheme (PPPT) as an additional storage overhead with reference to RPL-only.

\subsection{Simulation Results} \label{results}
Fig. \ref{graph:E_With_DutyCycling} shows network-wide energy usage in a duty-cycled RPL-based network with varying data rates (i.e., the number of packets sent per second) at different hop counts.
We may conclude that the suitable data rate for provenance-enabled RPL-based networks can be around 1 packet/20 seconds to 1 packet/30 seconds, whereas 1 packet/40 seconds can be considered as an ideal data rate. Moreover, the overall energy consumption by nodes in the network is loosely linked with the hop counts. 

Fig. \ref{graph:P_With_DutyCycling} shows average power consumption per-node against varying data rates. Here, per-node power decreases with the increase in data rate as fewer packets are being sent thereby utilizing less power. Hence, we can infer that per-node power consumption is more likely to be related to the data rate and less likely to be related with the hop count, similar to network-level energy consumption.
\begin{figure}[ht!]
\centerline{\includegraphics[width=2.5in]{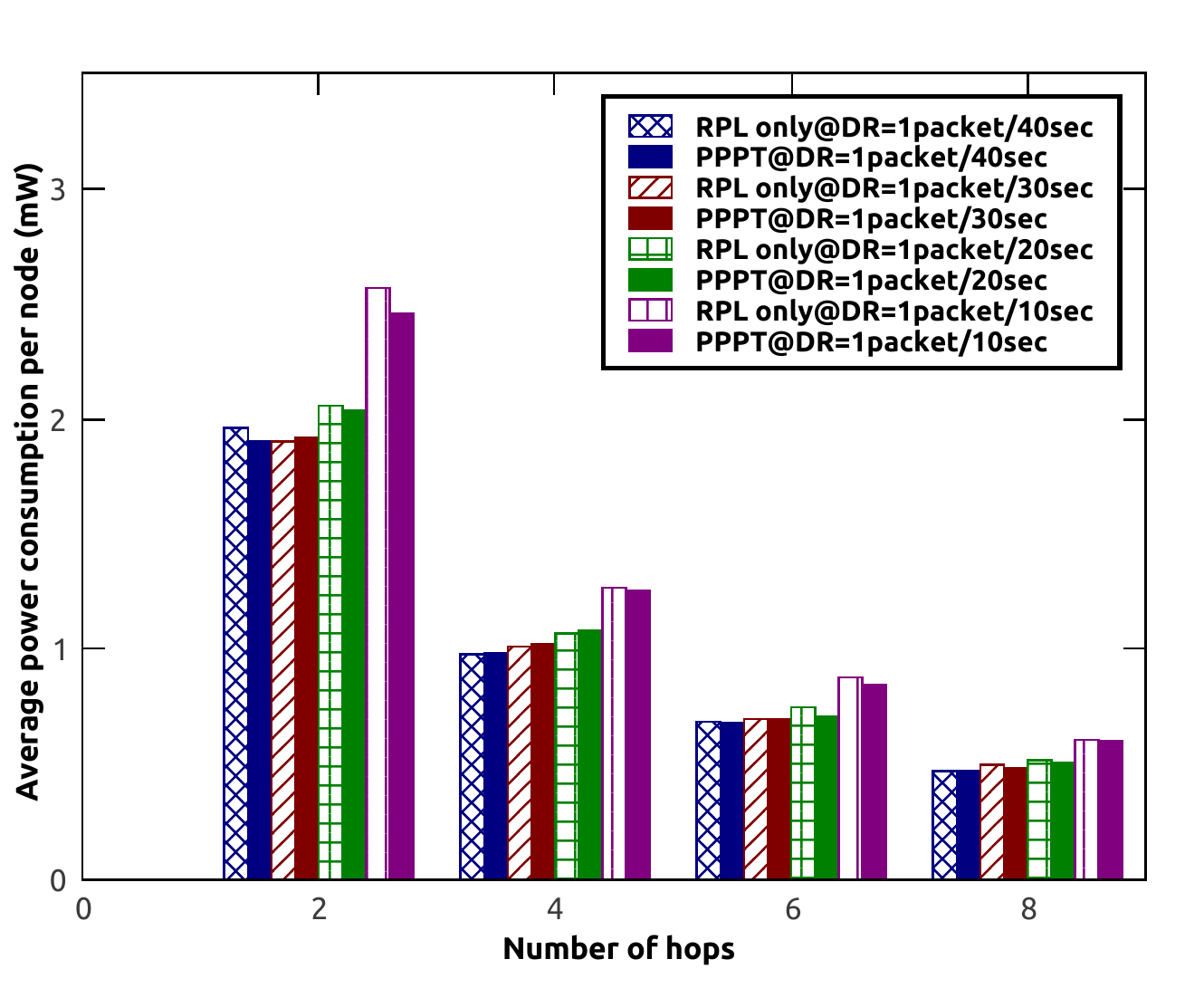}}
\caption{Comparison between \textit{RPL-only} and \textit{provenance-based RPL (PPPT)} showing average power consumption per node.}
\label{graph:comparison_RPL_PPPT}
\end{figure}
Fig. \ref{graph:comparison_RPL_PPPT} shows a comparison of node level power consumption in a duty-cycled network between RPL-only (without provenance) and PPPT (provenance-enabled RPL). By examining the results, we can see that power consumption is almost negligible in case of PPPT. In some cases, it is even less than RPL-only. This is because of reason that we are utilizing the available framework i.e., routing table for storing node-level information. 
\begin{figure}[t!]
\centerline{\includegraphics[width=2.5in]{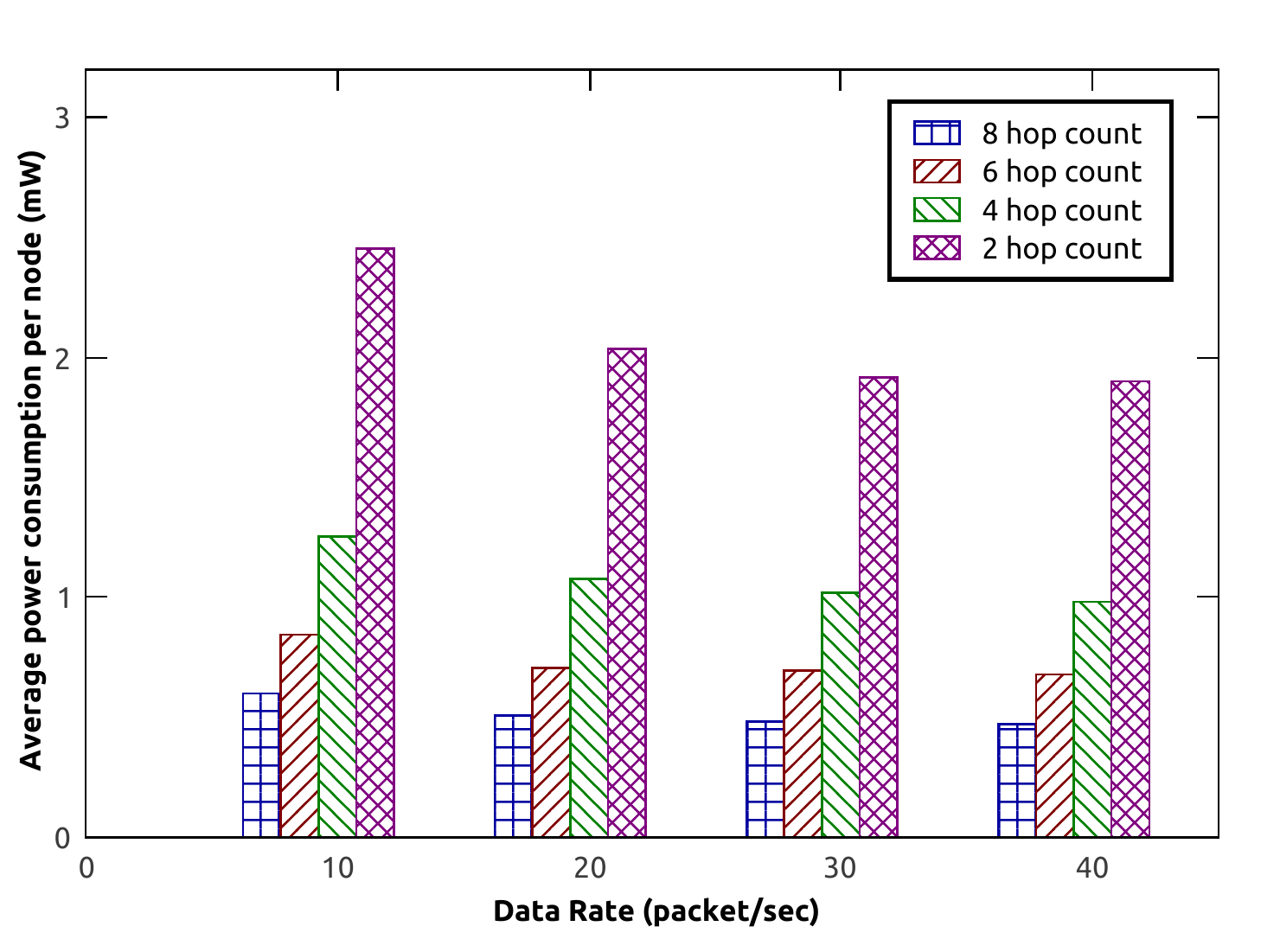}}
\caption{Average power consumption with duty cycling at different data rates.}
\label{graph:PC_vs_DR}
\end{figure}

\begin{figure}[t!]
\centerline{\includegraphics[width=2.5in]{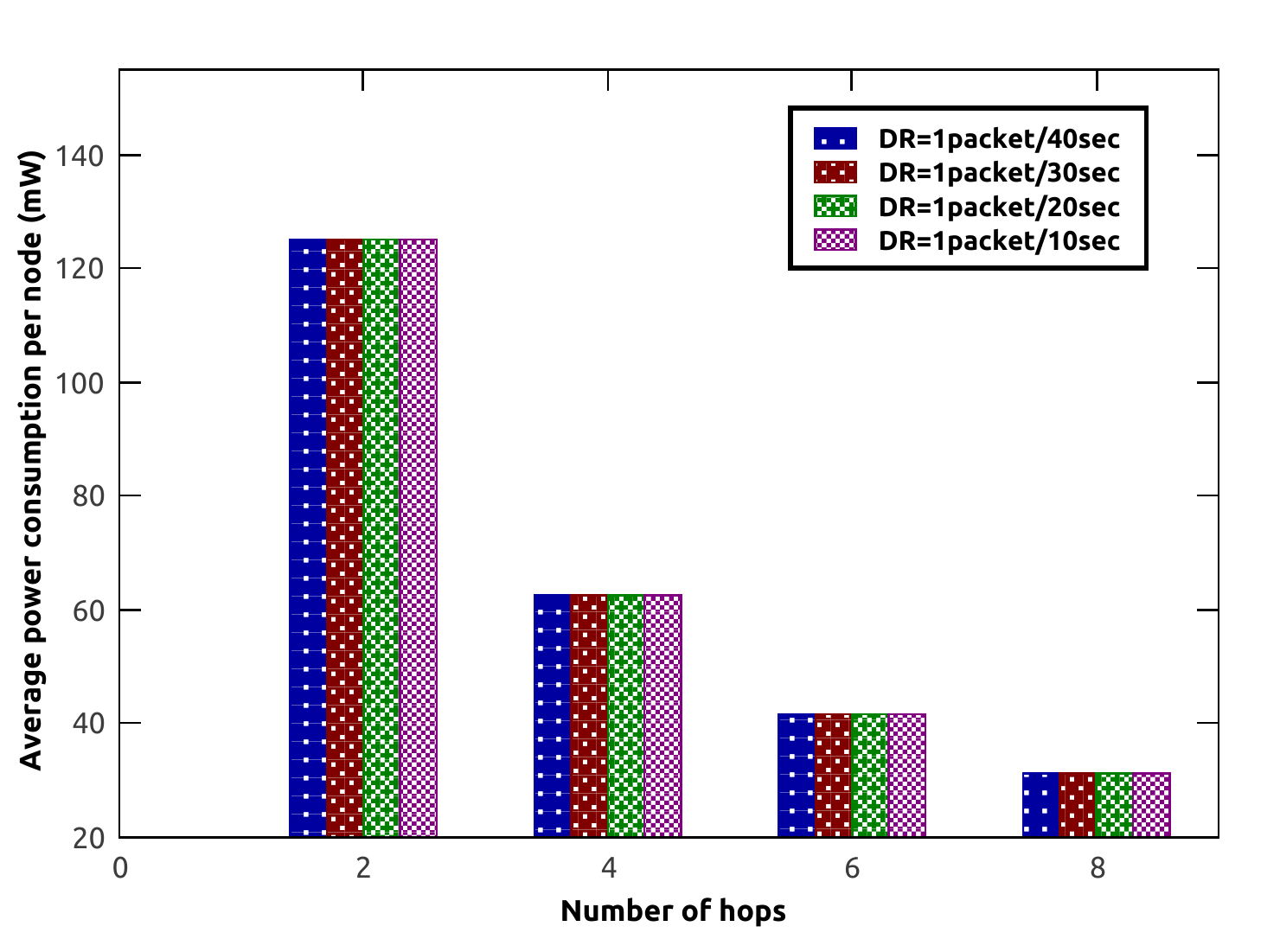}}
\caption{Average power consumption per-node without duty-cycling in a RPL-based provenance-enabled network.}
\label{graph:PC_withoutDC}
\end{figure}
\begin{figure*}[htbp]
\centering
\begin{subfigure}[b]{0.32\textwidth}
\centering
\includegraphics[width=2.5in, height=1.7in]{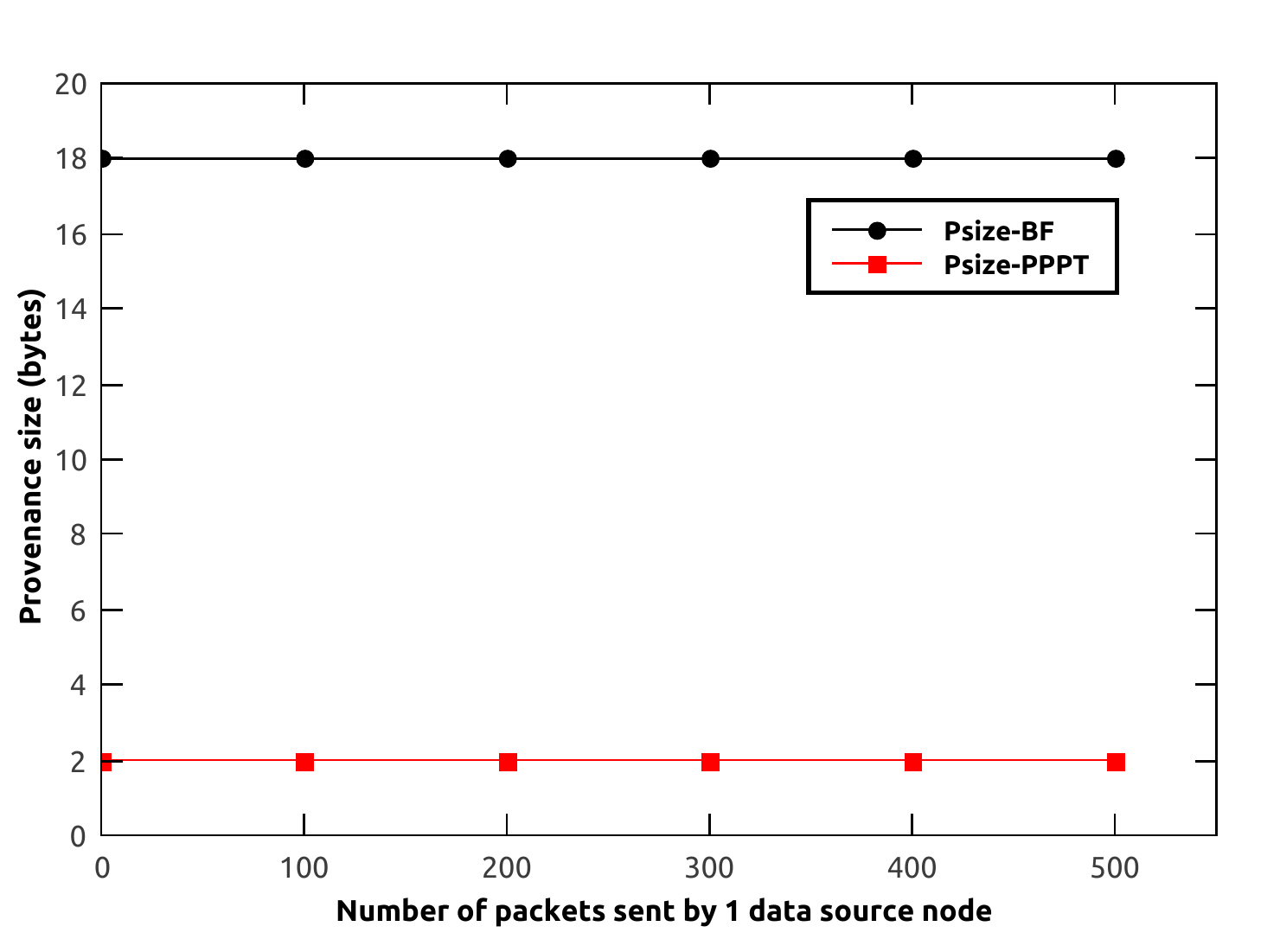}
\caption{}
\label{graph:Psize_1SN}
\end{subfigure}%
 \vspace{0.2cm}
\hspace*{\fill}
\begin{subfigure}[b]{0.32\textwidth}
\centering
\includegraphics[width=2.5in, height=1.7in]{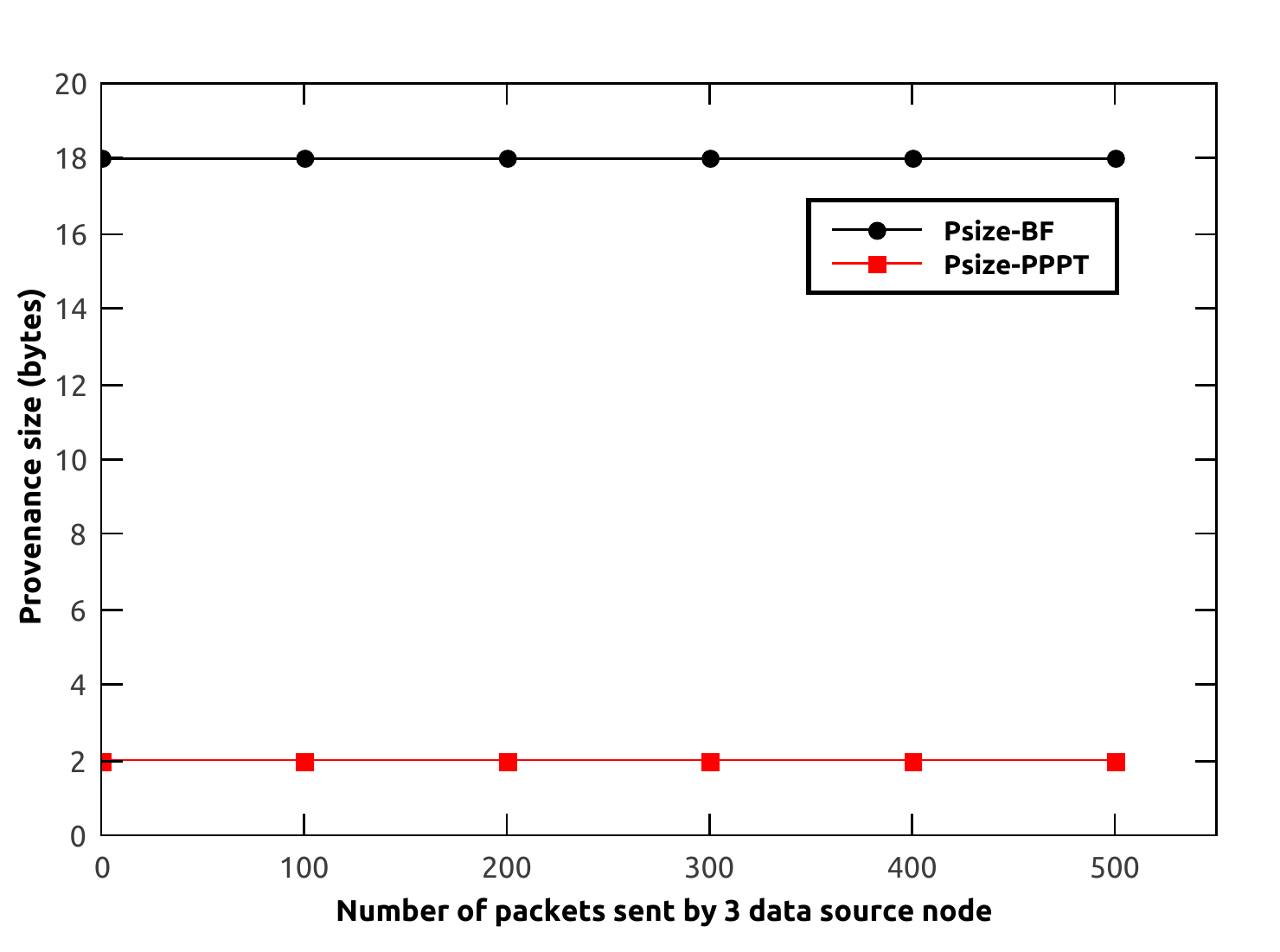}
\caption{}
 \label{graph:Psize_3SN}
\end{subfigure}
 \vspace{0.2cm}
\hspace*{\fill}
\begin{subfigure}[b]{0.32\textwidth}
\centering
\includegraphics[width=2.5in, height=1.7in]{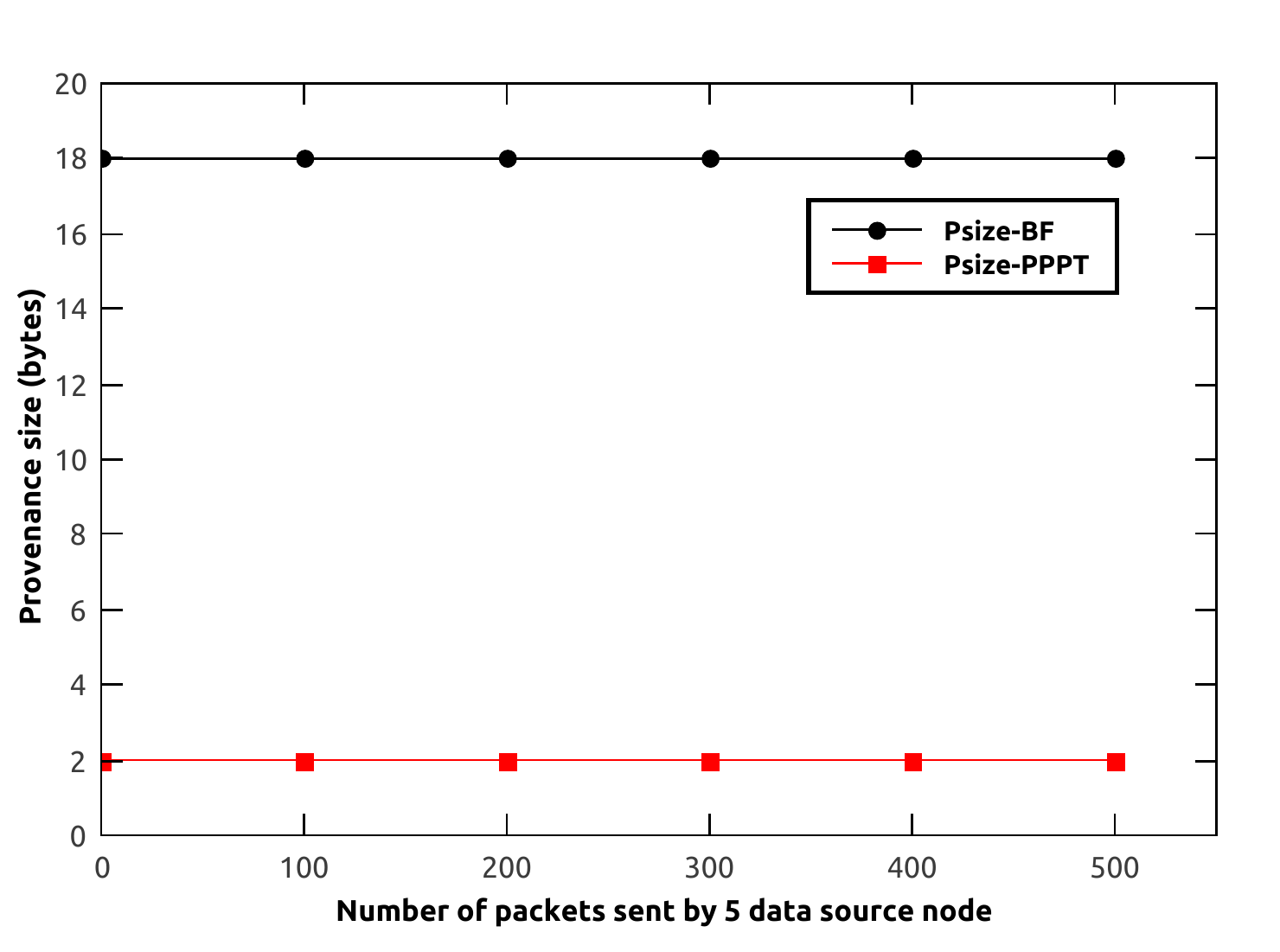}
\caption{}
 \label{graph:Psize_5SN}
\end{subfigure}
\centering
\caption{ Provenance size for (a) 1, (b) 3, and (c) 5 data source nodes. }
\label{graph:Psize_sourceNodes}
\end{figure*}

Fig. \ref{graph:PC_vs_DR} shows another perspective of the results shown in Fig. \ref{graph:P_With_DutyCycling}. Here, we can see that the average power consumption of per-node is more closely concerned with the data rate rather than the hop count. Hence, increase in the hop count may not affect much on per-node power consumption. However, it is evident from the result that as the data rate increases, the average power consumption per-node decreases. Fig. \ref{graph:PC_withoutDC} shows average power consumption per-node without duty-cycling in a RPL-based network. Here, we can see that as the number of hops are increasing, average power consumption per node is decreasing. This is because the nodes are operating without duty-cycling i.e., the radio is always on and nodes never sleep, hence power consumption is irrespective of the number of hops.
\begin{figure}[htbp]
\centerline{\includegraphics[width=2.5in]{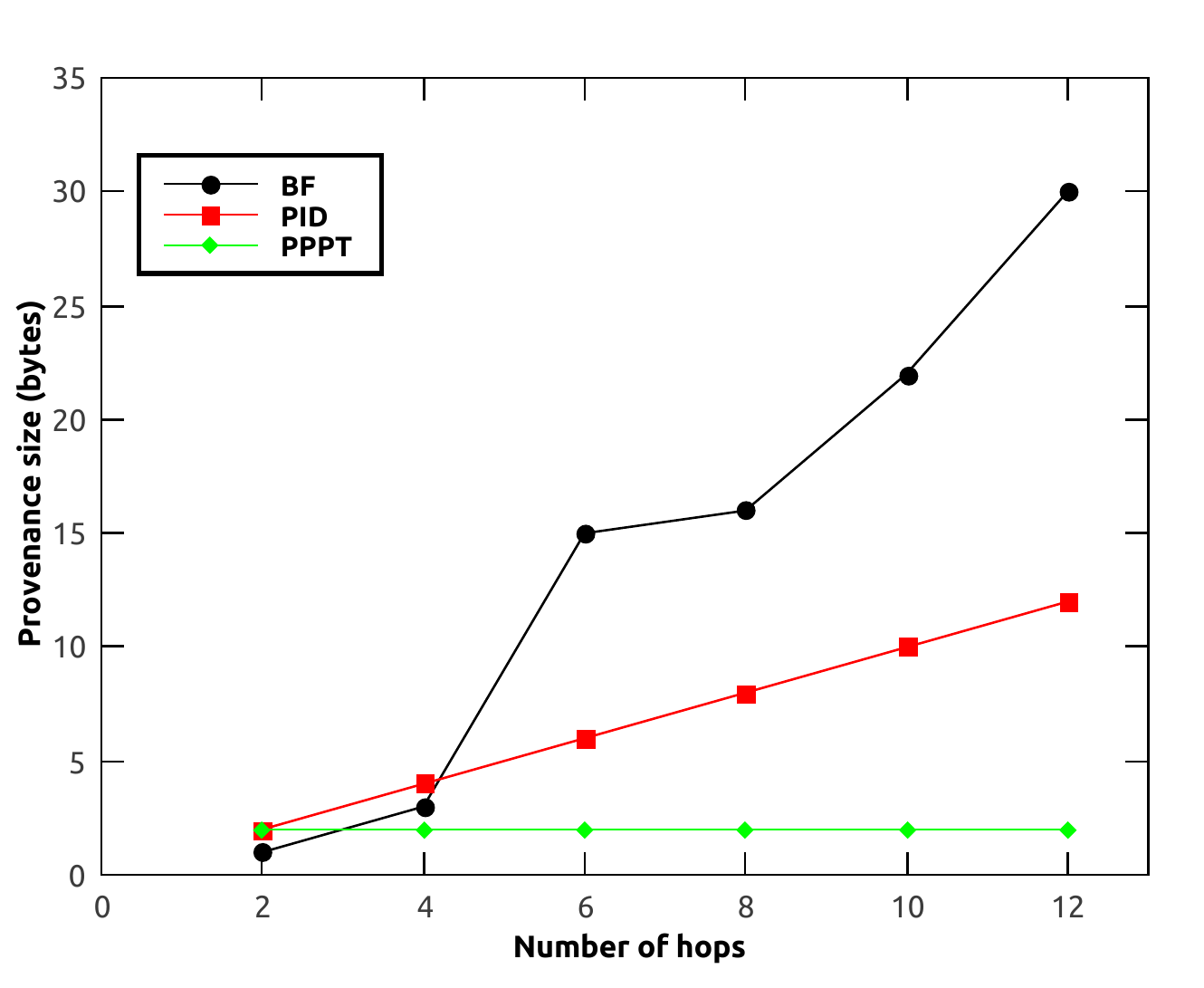}}
\caption{Comparison of provenance size among \textit{PPPT}, \textit{BF} \cite{sultana2013secure,sultana2015lightweight} and \textit{PID} \cite{suhail2018data} at different hop counts.}
\label{graph:Psize}
\end{figure}
Fig. \ref{graph:Psize} shows the comparison of proposed scheme with BF-based schemes \cite{sultana2013secure,sultana2015lightweight} and PID \cite{suhail2018data} in terms of provenance size ($P_{size}$) at different hop counts. In the case of PID, $P_{D}$ consists of the node IDs of all of the participating nodes that are responsible for carrying the data. Hence, $P_{size}$ increases with the increase in the number of hops. More specifically, $P_{size}$ (bytes) will be the same as that of hop count. However, in the case of PPPT, $P_{D}$ includes next destination node ID and current source node ID at each respective hop on the packet path. Therefore, $P_{size}$ remains constant i.e. 2 bytes (one byte for source node ID and one byte for destination node ID). 
We have also computed $P_{size}$ for the BF-based schemes that uses a fixed size BF to encode all the nodes on a packet’s path using a set of hash functions. $P_{size}$ for the BF scheme is determined by the size of the BF. Fig. \ref{graph:Psize_sourceNodes} shows that $P_{size}$ remains 2 bytes irrespective of number of packets sent by varying data source nodes. On the other hand, for BF scheme $P_{size}$ is much greater i.e., 18 bytes.

\begin{figure*}[!ht]
\centering
\begin{subfigure}[b]{0.48\textwidth}
\centering
\includegraphics[width=2.5in]{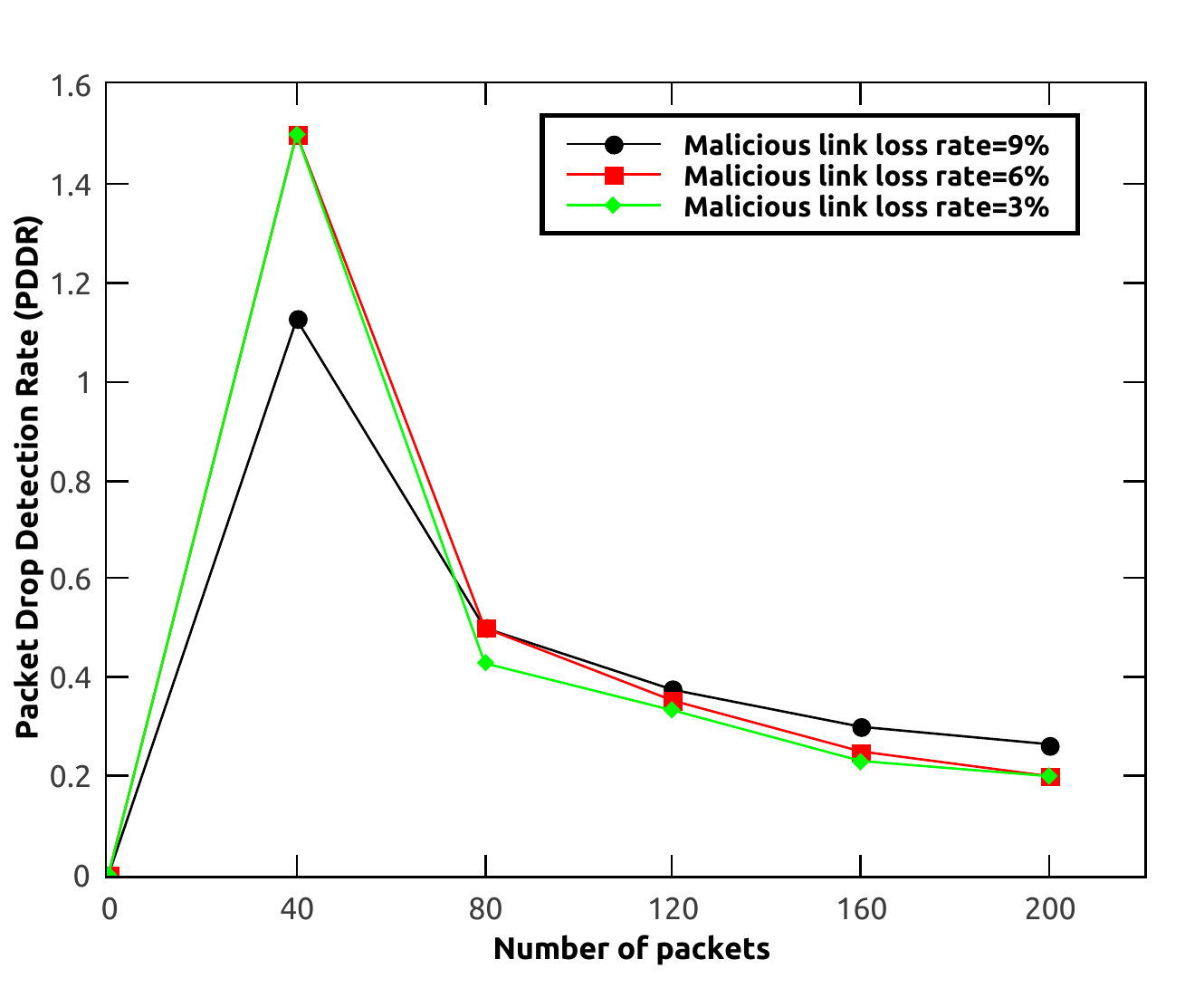}
\caption{}
\label{graph:PDDR_Packets}
\end{subfigure}%
\hspace*{\fill}
\begin{subfigure}[b]{0.48\textwidth}
\centering
\includegraphics[width=2.5in]{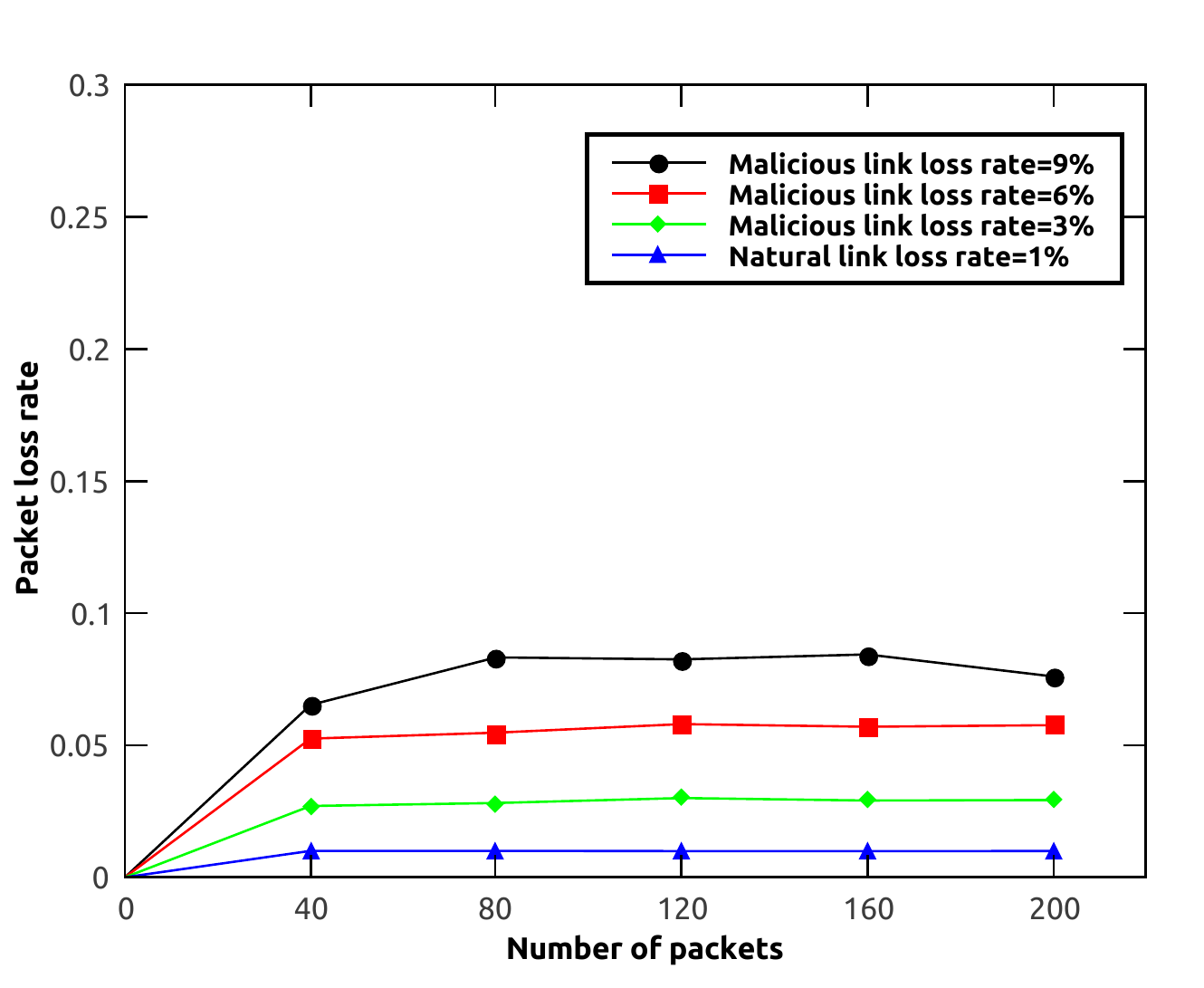}
\caption{}
 \label{graph:PacketLoss}
\end{subfigure}
\centering
\caption{(a) $PDDR$ over number of packets sent, (b) $PDDR$ for natural and malicious link loss rate.}
\end{figure*}

The result in Fig. \ref{graph:PDDR_Packets} shows that the higher the link loss rate is, the more the detection rate decreases. It is observed that as the link loss rate increases from 3\% to 9\%, the probability of consecutive packet drops also increases.
In order to detect packet drop attack, we perform a simulation in a 8-hop network and set node ID 3 as a malicious node. We also fix the natural link loss rate to 1\%, and vary the malicious link loss rate to 3\%, 6\%, and 9\%. Fig. \ref{graph:PacketLoss} shows the packet loss rate over the number of packets sent for a period of time. The malicious node drops data packets randomly, which eventually drops down the packet reception rate at the sink node. It can be concluded from the results that the packet loss rate for a benign network converges to the natural link loss rate. In contrast, due to the presence of malicious node, the packet loss rate gradually increases and ultimately reaches at a point that is  higher than the natural loss rate i.e., 1\%.

\begin{figure*}[!ht]
\centering
\begin{subfigure}[b]{0.48\textwidth}
\centering
\includegraphics[width=2.5in]{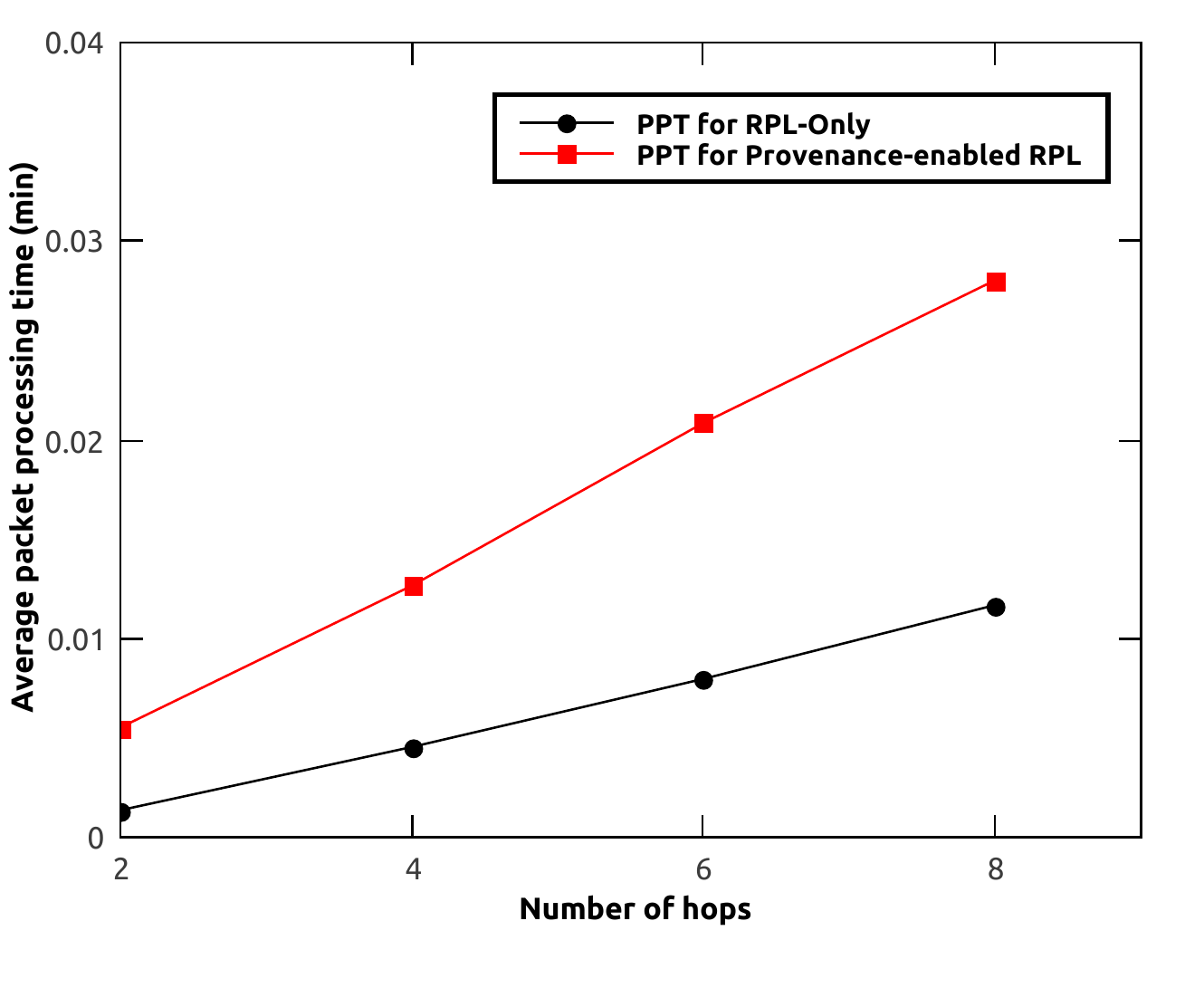}
\caption{}
\label{graph:PGT}
\end{subfigure}%
\hspace*{\fill}
\begin{subfigure}[b]{0.48\textwidth}
\centering
\includegraphics[width=2.5in]{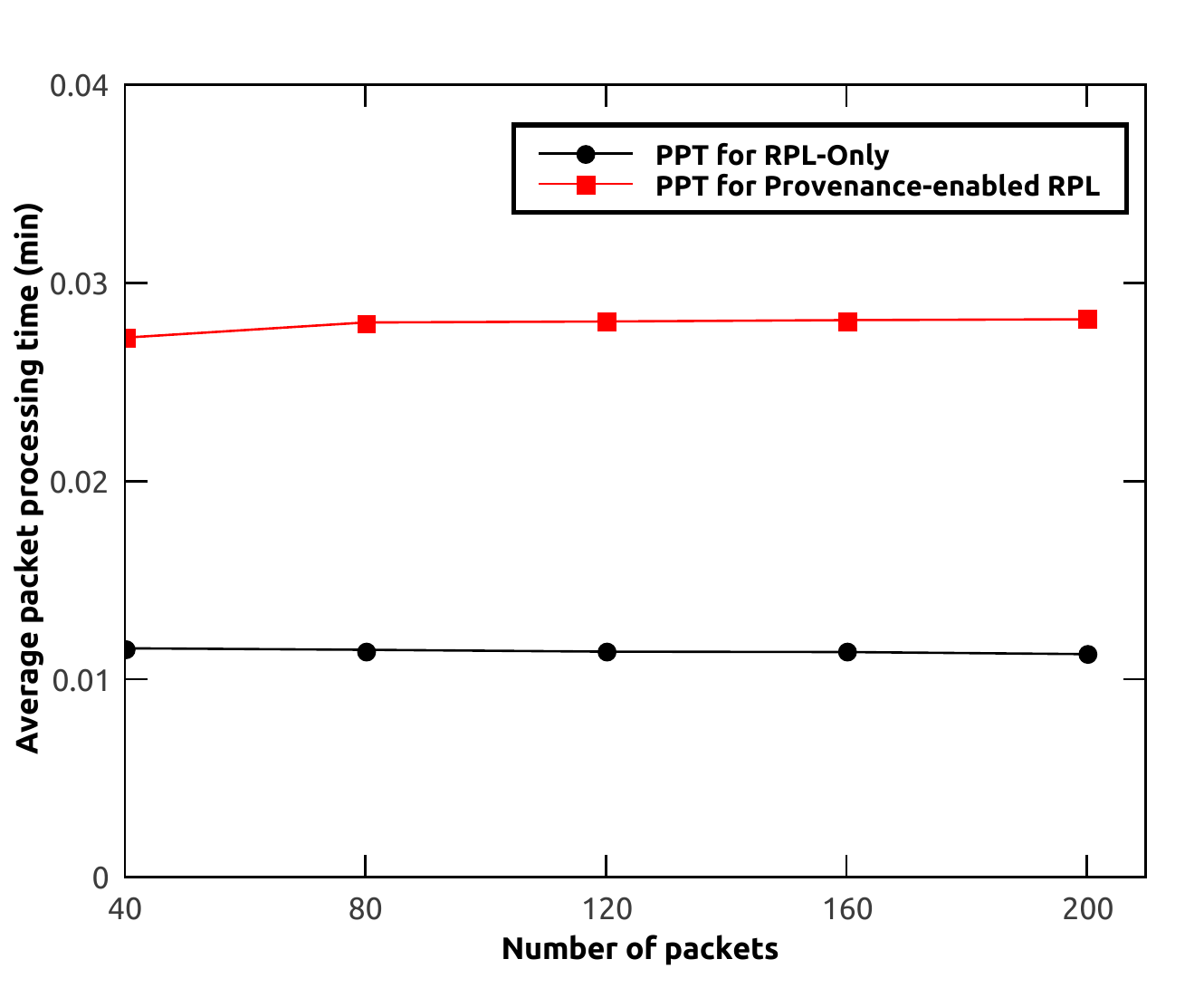}
\caption{}
\label{graph:PGT_packet}
\end{subfigure}
\centering
\caption{Comparison between \textit{RPL-only} and \textit{Provenance-based RPL (PPPT)} for average packet processing time with respect to (a) number of hops, (b) number of packets sent.}
\end{figure*}

Fig. \ref{graph:PGT} and Fig. \ref{graph:PGT_packet} shows the average packet processing time (PPT) (min) including other network processing delays. In Fig. \ref{graph:PGT} we can see that as the number of hops increases, the average PPT also increases, which is off-course natural in multi-hop network. The overall average PPT between RPL-only and Provenance-enabled RPL (PPPT) is nearly 0.006 and 0.017 min respectively, which is almost negligible. On the other hand, in Fig. \ref{graph:PGT_packet} as the number of sent packets increases, PPT almost remains the same. The required average processing time for Provenance-enabled RPL (PPPT) and RPL-only is 0.035 and 0.014 min respectively, which is again almost negligible. 
\section{Related Work} \label{relatedwork}
The interconnectivity of 6LoWPAN networks with the Internet raises serious security concerns, as resource-constrained things are globally accessible anywhere from the untrusted Internet. An attacker may attempt to disrupt the data generated by nodes either by inserting malicious nodes or by compromising benign nodes. Such forging or alteration of data produces catastrophic results, especially for applications relying on data for critical decision-making processes, risk assessment, and performance evaluation. To enable the data trustworthiness among IoT devices, provenance can be deployed. 

Extensive research has been carried out on network provenance. However, due to limited computational ability, energy constraints, and low bandwidth, the traditional provenance schemes, for instance, \cite{goodrich2008probabilistic,zhou2010efficient,zhou2011secure} cannot be applied to WSNs environment directly. Similarly, the concept of hop-by-hop tracing (or IP tracing) \cite{siddiqui2012hop} can be mapped to provenance, however, it is not well-suited to WSNs due to its large storage requirement. Therefore, existing provenance schemes developed for conventional
wired networks cannot be applied directly to WSNs due to both the resource-tightened nature of WSNs and the rapid provenance size increase \cite{wang2016provenance}.
\begin{table*}[t!]
  \centering
\caption{Quick overview of Lossy and Lossless Provenance schemes for WSNs } \label{tab:comparison}
  \begin{tabular}{|l|p{5cm}|p{5cm}|l| } 
   \hline
 \rowcolor{lightgray!30}
\textbf{Scheme} & \textbf{Methodology Used} &  \textbf{Weakness} & \textbf{Provenance Type} \\ [0.5ex] 
 \hline
\cite{shebaro2012demonstrating,sultana2013secure,sultana2015lightweight}  & Use bloom filter (BF) to record node IDs. & 
\begin{enumerate}[i)] 
\item Requires large sized BF to avoid high false positives. 
\item Only maintains node IDs as provenance information hence unable to construct packet path’s topology.
\end{enumerate} & lossy \\ 
\hline
\cite{DBLP:conf/wowmom/AlamF11} & Use probabilistic methods (including rank, prime and Rabin fingerprints) to incorporate the node IDs in the packet path. & \begin{enumerate}[i)]
\item Use of computationally intensive methods to encode provenance.
\item Frequent changes in data flow paths tends to scatter the provenance across several packets and thus incurs high decoding error rates.
\end{enumerate} & lossy \\ 
\hline
\cite{wang2016dictionary} & Maintains a dictionary to record packet path. & \begin{enumerate}[i)] 
\item Requires stable WSNs otherwise the packet may not reuse a packet path resulting in no compression at all.
\end{enumerate} & lossless \\
\hline
\cite{xu2018cluster,hussain2014secure}   & Use Arithmetic Coding to encode provenance as a coding interval. \cite{xu2018cluster} use global probabilities to encode provenance while \cite{hussain2014secure} use local probabilities to encode provenance. & \begin{enumerate}[i)]
    \item Requires training phase to assign the occurrence probability to each node.
    \item Requires transmitting two real numbers as coding interval which expands the provenance size.
    \item Requires the WSNs to stay stationary after deployment to avoid drastic modification of the occurrence frequencies and the probabilities assigned to the nodes.
\end{enumerate} & lossless \\
\hline
\end{tabular}
 \end{table*}
In sensor networks, \cite{shebaro2012demonstrating, sultana2015lightweight,sultana2013secure} proposed a provenance scheme that employ append-based data structures (bloom filter) to store provenance. In-packet bloom filter based provenance scheme works by embedding all nodes on a packet’s path in the BF using a set of hash functions. Upon packet retrieval, the BS retrieves the nodes on the path based on a certain false positive rate that is proportional to the size of the BF i.e., the larger the BF size, the lower the false positive rate. 

The authors of \cite{DBLP:conf/wowmom/AlamF11} proposed a probabilistic approach to encode the node identity into the provenance as an extension of IP traceback problem \cite{savage2000practical}. In this approach, each node makes an independent decision whether to embed its identity into the packet or not. Upon packet retrieval, the BS constructs provenance by exploiting partial path information collected from a number of received packets.

The main drawbacks associated with these \emph{lossy provenance schemes} are that (i) only nodes' IDs are recorded as the provenance information, (ii) increase in the provenance size with the number of nodes traversed, (iii) employ append-based data structures to store  provenance leading to prohibitive costs and high decoding error rates, (iv) high energy and bandwidth dissipation due to provenance size, and (v) provenance data loss.

Some \emph{lossless provenance schemes} have been designed to address the problem of increased provenance size in multi-hop networks by using data provenance compression. For instance, \cite{wang2016dictionary} proposed a dictionary-based provenance scheme by enclosing path index in the packet's path. In this scheme each node constructs a dictionary for a compressed packet’s path. The BS retrieves the path of the received packet by looking up the dictionary stored in it. \cite{hussain2014secure} proposed an arithmetic coding-based provenance scheme. This scheme assigns each node a global cumulative probability according to a node occurrence probability among all the used packet paths. \cite{xu2018cluster} presents an extension of \cite{hussain2014secure}. In this scheme, authors proposed an incremental cluster-based arithmetic coding scheme for encoding provenance in packet. Due to clustering-based mechanism the provenance on each layer can be encoded as an independent segment. Moreover, provenance is encoded through local probabilities rather than global probabilities thereby providing a higher compression rate. 

The main drawbacks associated with these techniques are that (i) unstable topology makes it hard to construct the packet path dictionaries leading to no compression at all, (ii) training phase is required in order to assign the occurrence probability to each node and, (iii) WSNs are required to stay stationary after deployment to avoid drastic modification of the occurrence frequencies and the probabilities assigned to the nodes.  Table \ref{tab:comparison} gives a quick overview of the methodology used and drawbacks associated with the existing lossy and lossless provenance schemes.
 
As the proposed PPPT scheme for RPL-based IoT devices is exploiting the existing features in RPL, hence, the shortcomings associated with lossy and lossless provenance schemes are feasible to resolve. For instance, the routing information is already maintained in routing table while sequence number is added against the routing information as a part of node-level information. Moreover, the current source node ID and next hop node ID are embedded as a part of system-level information, thereby reducing the provenance size to a constant size of 2 bytes. Similarly, the establishment of topology follows a proactive approach where the routing information is maintained in the routing table before sending packets.  

\section{Conclusion} \label{conclusion}
In this paper, we have proposed a provenance-enabled packet path tracing (PPPT) scheme for RPL-connected IoT devices to ensure the objective of data trustworthiness during packet traversal from source to destination. We have introduced node-level provenance by embedding sequence number against routing entry at the routing table of the respective forwarding node. In addition to node-level provenance, we have also added system-level provenance that captures the complete packet trace by including destination and source node ID pairs. The combination of both node and system level provenance verifies the provenance. Experimental results show that our scheme reduces the size of provenance and can save more energy and bandwidth with respect to BF-based schemes. Furthermore, the provenance generation time and average power consumption between RPL-only and provenance-based RPL is almost negligible. We like to introduce provenance scheme for the non-storing mode in RPL as a part of our future work.


%





\ifCLASSOPTIONcaptionsoff
  \newpage
\fi

\end{document}